\documentstyle[prd,aps,eqsecnum]{revtex}

\begin{document}
\title{Null dust in canonical gravity}  
\author{Ji\v{r}\'{\i} Bi\v{c}\'{a}k \footnote[1]{E-mail address:
bicak@mbox.troja.mff.cuni.cz}} 
\address{Department of Theoretical Physics,\\
Faculty of Mathematics and Physics, Charles University,\\
V Hole\v{s}ovi\v{c}k\'{a}ch 2, 180 00 Prague, Czech Republic}
\author{Karel V. Kucha\v r \footnote[2]{E-mail address:
 kuchar@mail.physics.utah.edu}} 
\address{Department of Physics, University of Utah,\\ 
   Salt Lake City, UT 84112, U.S.A.}
\maketitle 
\begin{abstract}
We present the Lagrangian and Hamiltonian framework which incorporates
null dust as a source into canonical gravity. Null dust is a
generalized Lagrangian system which is described by six Clebsch
potentials of its four--velocity Pfaff form. The Dirac--ADM
decomposition splits these into three canonical coordinates (the
comoving coordinates of the dust) and their conjugate momenta
(appropriate projections of four--velocity). Unlike ordinary dust of
massive particles, null dust therefore has three rather than four
degrees of freedom per space point. These are evolved by a Hamiltonian
which is a linear combination of energy and momentum densities of the
dust. The energy density is the norm of the momentum density with
respect to the spatial metric. The coupling to geometry is achieved by
adding these densities to the gravitational super--Hamiltonian and
supermomentum. This leads to appropriate Hamiltonian and momentum
constraints in the phase space of the system. The constraints can be
rewritten in two alternative forms in which they generate a true Lie
algebra. The Dirac constraint quantization of the system is formally
accomplished by imposing the new constraints as quantum operator
restrictions on state functionals. We compare the canonical schemes
for null and ordinary dust and emhasize their differences.
\end{abstract}
\nopagebreak[4]
\section{Introduction}

Null dust has been widely used as a simple matter source both in
classical and semiclassical gravity. Its equations of motion follow
directly from the conservation of the energy--momentum tensor. However,
the inclusion of null dust as a source into canonical gravity requires
careful identification of its own dynamical degrees of freedom. For
this purpose, one needs to construct a spacetime action depending on
appropriate Eulerian variables and bring it into canonical form by the
Dirac--ADM (Arnowitt, Deser and Misner) procedure. The coupling to
gravity, like that of other nonderivative systems, is then entirely
straightforward. The ordinary dust of massive particles was treated in
this manner by Brown and Kucha\v{r} \cite{B+K}. Our goal is to develop a
similar formalism for null dust.

The main application which we have in mind is minisuperspace and
midisuperspace quantization of canonical models which include null
dust as a source.  The specific models based on null dust are both
numerous and simple.  After the discovery of Vaidya's `radiating
Schwarzschild metric'
\cite{vai}, there were found many other, more general exact solutions
of Einstein's equations with null dust as a matter source.  Above all,
such models have recently been used to clarify the formation of naked
singularities during a spherical gravitational collapse, to describe
mass inflation inside black holes, and to model the formation and
Hawking evaporation of black holes. We briefly review these topics in
Appendix B. Our formalism is designed for studying such issues in
quantum rather than in classical or semiclassical contexts.

Null dust is intimately connected with the behavior of
zero--rest--mass fields in geometrical optics limit. The
energy--momentum tensor of such fields takes in that limit the form of
the energy-momentum tensor of null dust. One can then reinterpret some
exact solutions of Einstein's equations with null dust as spacetimes
produced by zero--rest--mass (in particular, electromagnetic)
fields. Careful studies of the high--frequency limit of the
gravitational radiation itself revealed that it also can be described
by the energy--momentum tensor of null dust. Moreover, which is
especially relevant for the present paper, such a connection can be
established at the level of a variational principle. All of this
indicates that null dust is much more closely related to fundamental
fields than ordinary dust formed by phenomenological massive
particles. We explain some of these connections in Appendix A.

We start our exposition by reviewing how the dynamics of incoherent
dust follows from the conservation law of the energy--momentum tensor
(Section 2). This enables us to pinpoint at the very beginning the
main difference between ordinary dust and null dust: The normalization
of timelike four--velocity selects its parametrization by proper time,
the null normalization of lightlike four--velocity leaves its
parametrization arbitrary. This is of paramount importance both for
the Lagrangian and Hamiltonian descriptions of null dust.

Since null geodesics are somewhat less intuitive than timelike
geodesics, we briefly summarize the basic properties of null
congruences in Section 3. We explain how to obtain an affine
parametrization of such congruences, but stress its essential
ambiguity which prevents the unique separation of mass distribution of
null dust particles from their four--velocity. When it comes to
producing the gravitational field, the mass distribution can simply be
reabsorbed into four--velocity, which is the reason why it does not
naturally occur as a separate variable either in the Lagrangian or
Hamiltonian frameworks. We formulate a spacetime variational principle
from which the null geodesic equations of motion follow in Section
4. The variational principle for null dust is quite similar to the
variational principle for ordinary dust given by Brown and Kucha\v{r}
\cite{B+K}, but there are several characteristic differences.  The most
important one was already mentioned: Null worldlines have no natural
parametrization and hence the null velocity appears in the variational
principle as the Pfaff form of six scalars (the Clebsch potentials)
rather than seven scalars which characterize the timelike
velocity. Consequences of this distinction can be traced throughout
the whole formalism. In Appendix C we illustrate on explicit examples
the decomposition of null covector fields into Clebsch potentials
which is a prerequisite of our variational principle. In Section 5 we
show that any solution of the Euler equations of the variational
principle provides enough building blocks to reconstruct an affinely
parametrized four--velocity $k^{\alpha}$ and the associated mass
distribution $M$. We also discuss other special parametrizations. In
Section 6, we cast our covariant spacetime action into Hamiltonian
form by following the Dirac--ADM algorithm. The details of this
process are substantially different from the steps which need to be
taken for ordinary dust. We express the energy and momentum densities
of null dust in terms of appropriate canonical variables. The energy
density turns out to be the norm of the momentum density with respect
to the spatial metric. It transpires that null dust has only three
degrees of freedom per space point, one less than ordinary dust which
has four. The missing degree of freedom is a privileged scalar
parameter (like proper time) along lightlike geodesics. The missing
canonically conjugate momentum is the mass distribution which has been
reabsorbed into the four--velocity form. We conclude this section by
writing down the standard Hamiltonian and momentum constraints for
geometry coupled to null dust. In Section 7, we rewrite these
constraints in two alternative forms in which they generate a true Lie
algebra.  In this process, the Hamiltonian constraint is replaced by
alternative constraints which contain only geometric variables. This
feature of the constraint is related to a Rainich--type `already
unified theory' for geometry coupled to null dust. In Section 8, we
show how one can formally impose the new form of constraints as
quantum operator restrictions on state functionals. The outcome of
this procedure is a single functional differential equation for
physical state functionals $\bf{\Psi} [\bf{g}]$ which depend solely on
the spatial metric $\bf{g}$ in the dust frame. In the final Section 9,
we compare the canonical formalism and the ensuing quantum theory for
null dust with those for ordinary dust and emphasize their
differences.

Our conventions follow those of Misner, Thorne and Wheeler \cite{MTW},
except for our choice of units which are such that $16\pi G=1=c$.

\section{Dust as a source of gravity}

Incoherent dust is one of the simplest phenomenological sources of
gravity in general relativity. Its energy--momentum tensor
\begin{equation}
T^{\alpha \beta} = M U^{\alpha}  U^{\beta} ,
\end{equation}
\begin{equation}
U^{\alpha}  U_{\alpha} = -1 \, ,
\end{equation}
curves the spacetime according to Einstein's law of gravitation
\begin{equation}
G^{\alpha \beta} := R^{\alpha \beta} - \frac{1}{2} R\gamma^{\alpha \beta}
=  \frac{1}{2} T^{\alpha \beta}\,.
\end{equation}

Dust is described by the four--velocity $U^{\alpha}$ of its particles
and the (rest) mass density $M$ of their distribution. The equations
of motion of the dust are entirely contained in the energy--momentum
conservation law
\begin{equation}
\nabla_{\beta} T^{\alpha \beta} = 0
\end{equation}
which follows from the Einstein law (2.3) through the Bianchi
identities.  The structure (2.1) of the energy--momentum tensor allows
us to write Eq. (2.4) in the form
\begin{equation}
M U^{\beta} \nabla_{\beta} U^{\alpha} + \nabla_{\beta}(M U^{\beta})
U^{\alpha} = 0 \,.
\end{equation}
One sees that
\begin{equation}
M U^{\beta} \nabla_{\beta} U^{\alpha} \propto U^{\alpha} ,
\end{equation}
i.e., that the dust particles move along geodesics. The normalization
(2.2) of the four--velocity tells us that the particle worldlines are
parametrized by proper time. When one multiplies the geodesic equation
by $ U_{\alpha}\,$, the normalization condition (2.2) implies the rest
mass conservation
\begin{equation}
\nabla_{\beta} (M U^{\beta}) = 0 \,.
\end{equation}
By using Eq. (2.7) back in Eq. (2.5) one learns that proper time
is an affine parameter:
\begin{equation}
U^{\beta} \nabla_{\beta} U^{\alpha} = 0 \,.
\end{equation}
These facts describe in full detail the motion of ordinary dust of
massive particles.

Null dust has the same energy--momentum tensor (2.1) as ordinary dust,
but its particles are assumed to follow lightlike worldlines:
\begin{equation}
U^{\alpha}  U_{\alpha} = 0 \, .
\end{equation}
The energy--momentum conservation law (2.4) still implies that those
worldlines are geodesics, Eqs. (2.5)--(2.6). However, the null
normalization (2.9) no longer enforces either the conservation law
(2.7) or affine parametrization (2.8).

For ordinary dust, the decomposition of the energy--momentum tensor
into the mass density $M$ and four--velocity $ U_{\alpha}$ is unique
due to the timelike normalization (2.2). For null dust, the lightlike
normalization (2.9) is preserved by an arbitrary scaling of $
U_{\alpha}\,$:
\begin{equation}
\overline{U}^{\alpha} = \Lambda U^{\alpha}, \;\; \Lambda > 0 \,. 
\end{equation}
(The limitation $ \Lambda > 0 $ is needed to preserve the
future--pointing orientation of the worldlines.) By simultaneously
rescaling the scalar $M\,$,
\begin{equation}
\overline{M} = {\Lambda}^{-2} M \,, 
\end{equation}
one preserves the form (2.1) of the energy--momentum tensor. This shows
that the decomposition of $T^{\alpha \beta}$ into $M$ and $U^{\alpha}$
is arbitrary. In particular, by taking $\Lambda = M^{1/2}\,$, one can
eliminate the scalar $M$ altogether and write $T^{\alpha \beta}$
entirely in terms of a single null vector
\begin{equation}
l^{\alpha} := M^{1/2} U^{\alpha} 
\end{equation}
as
\begin{equation}
T^{\alpha \beta} = l^{\alpha}  l^{\beta}.
\end{equation}
In terms of $l^{\alpha}$, the geodesic equation (2.5) takes the form
\begin{equation}
l^{\beta} \nabla_{\beta} l^{\alpha} + ( \nabla_{\beta} l^{\beta})
l^{\alpha} = 0 \,.
\end{equation}
We shall see later that this choice maximally simplifies the form of
the null dust action and its canonical decomposition.

\section{Null geodesic congruences}
In a region of spacetime, $\cal M$, which is filled by dust whose
worldlines do not intersect, the vector field $U^{\alpha}$ defines a
line congruence $\cal S$. This congruence can be viewed as an abstract
three--dimensional space, the `dust space', whose points are the
individual worldlines. The worldlines $z\in \cal S$ can be locally
labeled by three parameters $z^{k}(z)$ which introduce a coordinate
chart in $\cal S\,$. We shall use the indices $i,\,j,\,k$ from the
middle of the Latin alphabet to denote the components of the objects
in $\cal S\,$; they take the values $ 1,\, 2,\, 3\,$. (A global
standpoint replacing this local description is discussed in
\cite{B+K}.)

Through each event of the region there passes one and only one
worldline. One can uniquely assign to each event $y$ the labels
$z^{k}$ of that worldline:
\begin{equation}
z^{k} = Z^{k}(y) \,.
\end{equation}
Our interpretation of the scalar fields $ Z^{k}(y)$ presupposes that
their values $z^{k}$ constitute a good chart in $\cal S \,$.
Therefore, the three gradients ${Z^k}_{,\alpha}$ must be three
linearly independent covectors:\footnote[1]{The contravariant tensor
density $\delta ^{\alpha \beta \gamma \delta}$ of weight $1$ is the
alternating symbol in $\cal M\,$. The covariant tensor density $\delta
_{ijk}$ of weight $-1$ is the alternating symbol in $\cal S\,$. The
Levi--Civita pseudotensor in $\cal M$ is denoted by $\epsilon ^{\alpha
\beta \gamma \delta}$.}
\begin{equation}
U^{\alpha} \propto \frac{1}{3!}\, \delta ^{\alpha \beta \gamma \delta}
\,{Z^i}_{,\beta} {Z^j}_{,\gamma} {Z^k}_{,\delta}\, \delta_{ijk} \neq
 0 \,.
\end{equation}
Parametrize the curves of $\cal S$ by a parameter $u$ whose rate of
change judged by the size of $U^{\alpha}$ is unity. In other words, if
\begin{equation}
u = U(y)
\end{equation}
is the value of $u$ on the curve of $\cal S$ which passes through
the event $y$, it holds that
\begin{equation}
U^{\alpha}\nabla_{\alpha}U = 1 \,.
\end{equation}
Equations (3.2) and (3.4) imply that $Z^{K}=(U,\,Z^{k})$ are four
independent functions of spacetime coordinates $y^{\alpha}$:
\begin{equation}
\det({Z^K}_{,\alpha})\, =  \frac{1}{3!} \delta ^{\alpha \beta
\gamma \delta}\, U_{,\alpha} {Z^i}_{,\beta} {Z^j}_{,\gamma}
 {Z^k}_{,\delta}\, \delta_{ijk} \neq  0 \,.
\end{equation}
The mapping $\,Z\, : \, {\cal M} \, \rightarrow \,
{\hbox{$I$\kern-3.8pt $R$}} \times {\cal S}$ given locally by
Eqs. (3.3) and (3.1) can thus be inverted into the mapping $\,\Upsilon
\, : \, {\hbox{$I$\kern-3.8pt $R$}} \times {\cal S} \, \rightarrow \,
\cal M$ given locally by
\begin{equation}
y^{\alpha}=\Upsilon^{\alpha} (u, z^{k})\,.
\end{equation}
Here, $z^{k}$ distinguish different curves of the congruence $\cal S$
and $u$ specifies the point on a given curve. The four vectors
\begin{equation}
U^{\alpha} = {\Upsilon ^{\alpha}}_{,u}\,, \;\;\; Z^{\alpha}_{k} 
= {\Upsilon ^{\alpha}}_{,k}
\end{equation}
form a basis in $T \cal M$ dual to the cobasis
\begin{equation}
{Z^K}_{,\alpha} =(U_{,\alpha}\,,\, {Z^k}_{,\alpha})
\end{equation}
in $T^{*} \cal M\,$. The basis (3.7) and cobasis (3.8) satisfy the
standard orthonormality and completeness relations. In particular
\begin{equation}
Z^{\alpha}_{i} {Z^k}_{,\alpha} = \delta^{k}_{i} \,.
\end{equation}

So far, everything applies equally well both to timelike and null
congruences. Brown and Kucha\v{r} \cite{B+K} specialized the formalism
to timelike congruences and applied it to Lagrangian description of
ordinary dust. In this paper, we first briefly recapitulate how to
specialize the formalism to null geodesic congruences (see,
e.g., \cite{sachs} and \cite{P+R} for more details) and then use it for
Lagrangian description of null dust.

A geodesic null congruence $U^{\alpha}$ must satisfy the geodesic
condition (2.6) and the null condition (2.9). These conditions still
hold when the vector field $U^{\alpha}$ is scaled by an arbitrary
factor, Eq. (2.10). Instead of using that scaling for eliminating $M$
from the energy--momentum tensor, Eqs. (2.11)--(2.13), one can use it
for enforcing affine parametrization. In terms of $l^{\alpha}$, the
geodesic equation takes the form (2.14). Unless $l^{\alpha}$ happens
to be divergencefree, it is not affinely parametrized. Let us first
show that there exists a positive scaling factor
\begin{equation}
\Lambda (y) = {\rm e}^{\lambda (y)}
\end{equation}
such that
\begin{equation}
\nabla_{\beta} (\Lambda ^{-1} l^{\beta}) = 0 \,.
\end{equation}
The condition (3.11) amounts to a linear inhomogeneous equation
\begin{equation}
l^{\beta} \nabla_{\beta} \lambda = \nabla_{\beta} l^{\beta} 
\end{equation}
for $\lambda \,$.
In the adapted coordinates $u, z^{k}$, Eq. (3.12) assumes the form
\begin{equation}
\frac {\partial \lambda (u, z)}{\partial u}  = (\nabla_{\beta}
 l^{\beta}) (u, z) \,.
\end{equation}
Its general solution
\begin{equation}
\lambda (u, z) = \int ^{u}_{0} d u \, (\nabla_{\beta} l^{\beta})
(u, z) \, + \lambda_{0}(z)
\end{equation}
depends on an arbitrary function $ \lambda_{0}(z) $ of $z\,$.
By writing Eqs. (2.13)--(2.14) and (3.11) in terms of the new variables
\begin{equation}
k^{\alpha} \,\mbox{:=}\, \Lambda l^{\alpha} \;\; {\rm and} \;\;
M \,\mbox{:=}\, \Lambda ^{-2}
\end{equation}
one learns that the vector field $k^{\alpha}$ is affinely parametrized,
\begin{equation}
k^{\beta} \nabla_{\beta} k^{\alpha} = 0 \,,
\end{equation}
the mass distribution $M$ satisfies the continuity equation
\begin{equation}
\nabla_{\beta} (M k^{\beta}) = 0 \,,
\end{equation}
and the energy--momentum tensor takes the form
\begin{equation}
T^{\alpha \beta} = M k^{\alpha} k^{\beta}.
\end{equation}

The affine parameter $v$ is a monotonically increasing function of the
old parameter $u\,$:
\begin{equation}
v(u, z) = \int ^{u}_{0} d u \, \Lambda ^{-1}(u, z) \, + v_{0}(z) \,.
\end{equation}
When we define a new mapping $\,\Upsilon _{\rm AFF}^{\alpha} \, : \,
{\hbox{$I$\kern-3.8pt $R$}} \times {\cal S} \; \rightarrow \; \cal M$ by
\begin{equation}
\Upsilon _{\rm AFF}^{\alpha}(v,z)\, \mbox{:=}\, \Upsilon^{\alpha}
 {\bf (} u(v,z),u {\bf )}
\end{equation}
we obtain
\begin{equation}
k^{\alpha} = \frac{\partial \Upsilon _{\rm AFF}^{\alpha}(v,z)}
{\partial v} \,.
\end{equation}

The affine parametrization (3.19) depends on two arbitrary functions,
$\lambda_{0}(z)$  and $v_{0}(z)\,$. This means that along each geodesic
the affine parameter is determined only up to a linear transformation
\begin{equation}
\bar{v} = \Lambda_{0}^{-1}(z)\,v + \bar{v}_{0}\,, \;\; \Lambda_{0}(z)>0\,.
\end{equation}
When we change the affine parameter by Eq. (3.22), the null vector field
$k^{\alpha}(y)$ is scaled into
\begin{equation}
\bar{k}^{\alpha}(y) = \Lambda_{0}{\bf (}Z(y){\bf )}\, k^{\alpha}(y) \,.
\end{equation}
The affinely parametrized null vector field $ k^{\alpha}(y)$ is thus
determined only up to an arbitrary positive multiplicative factor
$\Lambda_{0}$ which is a function of comoving coordinates
$z^{k}=Z^{k}(y)\,$.

A congruence of affinely parametrized null geodesics is characterized
by its twist (or rotation) $\omega$, expansion $\theta$, and shear
$\sigma$. The corresponding scalars are given by
\begin{eqnarray}
\omega &=& \left(\frac{1}{2}\nabla _{[\alpha}k_{\beta]}
 \nabla ^{\alpha}k^{\beta}
\right)^{1/2}\!,\\
\theta &=& \frac{1}{2}\left( \nabla _{\alpha}k^{\alpha}\right)\,,\\
|\sigma | &=& \left(\frac{1}{2} \nabla _{(\alpha}k_{\beta)}\nabla
^{\alpha}k^{\beta}-\theta^2  \right)^{1/2}\!,
\end{eqnarray}
where the square (round) brackets around indices denote
antisymmetrization (symmetrization).
The twist can also be determined from the relation
\begin{equation}
\omega k^{\alpha}=\frac{1}{2}
\epsilon^{\alpha\beta\gamma\delta}k_{\beta}\nabla_{\delta}k_{\gamma}\,.
\end{equation}
If $k_\alpha$ is proportional to a gradient,
\begin{equation}
k_{\alpha}(y) = \phi (y)\psi_{,\alpha}(y) \,,
\end{equation}
the geodesics of ${\cal S}$ form null hypersurfaces $\psi = const$ to
which $k^\alpha$ is orthogonal. A null geodesic congruence is
hypersurface orthogonal if and only if it has a vanishing twist:
$\omega = 0\,$.

Under the change (3.23) of affine parametrization, the rotation,
expansion and shear all scale by the same factor:
\begin{equation}
\bar{\omega}= \Lambda_{0} \omega \,, \; \;
\bar{\theta}= \Lambda_{0} \theta \,, \; \;
|\bar{\sigma}| = \Lambda_{0} |\sigma| \,.
\end{equation}
Also, by using Eqs. (3.10)--(3.12) and (3.15),
one can reexpress them in terms of $l^{\alpha}$ and its derivatives,
and of the undifferentiated scaling factor (3.10), (3.14):
\begin{eqnarray}
\omega &=& \Lambda \left(\frac{1}{2}\nabla _{[\alpha}l_{\beta]} \nabla
^{\alpha}l^{\beta} +\frac{1}{4} (\nabla_{\alpha} l^{\alpha})^{2}
 \right)^{1/2},\\
\theta &=& \Lambda \, \left( \nabla _{\alpha}l^{\alpha}\right) \,,\\
|\sigma | &=& \Lambda \left(\frac{1}{2} \nabla
_{(\alpha}l_{\beta)}\nabla ^{\alpha} l^{\beta} - \frac{5}{4}
(\nabla_{\alpha} l^{\alpha})^2 \right)^{1/2}\! .
\end{eqnarray}
The scalars (3.30)--(3.32) allow us to introduce other special
parametrizations of null geodesic congruences. Considerations about
the rate of expansion of a shadow image (see, e.g., \cite{sachs} and
\cite{P+R}) lead to the concept of luminosity distance $\Lambda$. This
is defined as any solution of the equation
\begin{equation}
\frac{1}{\Lambda} \frac{d \Lambda}{d v} = \theta \,,
\end{equation}
where $v$ is an affine parameter and $\theta$ (assumed to be
nonvanishing) is the expansion (3.25). Equations (3.15) and (3.23)
ensure that the luminosity distance is identical with the scaling
factor (3.10), (3.12). The luminosity distance played a prominent role
in several classical works in radiation theory \cite{bmb} and in
cosmology \cite{kr+sa}. The mass distribution $M$ introduced by
Eq. (3.15) is the inverse square $M=\Lambda^{-2}$ of the luminosity
distance. The parallax distance $p\,$,
\begin{equation}
p = \theta^{-1} \,,
\end{equation}
is also occasionally useful.
\section{Spacetime action and the Euler equations}
We describe null dust by six spacetime scalars $Z^k$, $W_k\,$. The
interpretation of our state variables $Z^k$, $W_k$ emerges from the
form of the action and the resulting equations of motion. We shall see
that $Z^k$ are comoving coordinates of null dust particles. By
specifying the values $z^k$ of the scalars $Z^{k}(y)$, we choose a
particular null geodesic of the congruence ${\cal S}$. The three
gradients ${Z^k}_{,\alpha}$ are assumed to be three independent
covectors. We shall see later that none of them can be timelike.

The four--velocity covector $l_\alpha$ of a lightlike particle is given
by its components $W_k$ in the cobasis ${Z^k}_{,\alpha} \,$:
\begin{equation}
l_{\alpha}=W_k {Z^k}_{,\alpha}\,.
\end{equation}
This relation expresses the one--form $l = l_\alpha dy^\alpha$ as 
a Pfaff form
\begin{equation}
l=W_k dZ^k
\end{equation}
of six scalar fields $Z^k$ and $W_k\,$. According to Pfaff's theorem
\cite{cara}, four scalar potentials $A, B, C, D$ are sufficient to
describe an arbitrary covector in a four--dimensional space:
\begin{equation}
l_{\alpha}=AB_{,\alpha}+CD_{,\alpha}\,.
\end{equation}
However, the representation of $l_\alpha$ by six potentials $W_k$,
$Z^k$ is more useful because it has a clear physical interpretation
\cite{FluidS}.

The null dust action
\begin{equation}
S^{\rm ND}[Z^k,W_k\,;\,\gamma_{\alpha\beta}]
=\int_{}^{}d^4y \, L^{\rm ND}(y)
\end{equation}
is a functional of our six state variables, and of the metric
$\gamma_{\alpha \beta}\,$. The Lagrangian density $L^{\rm ND}$ is
taken in the form
\begin{equation}
L^{\rm ND}=-\frac{1}{2}|\gamma|^{1/2}
 \gamma^{\alpha\beta}l_{\alpha}l_{\beta}\,,
\end{equation}
where $l_\alpha$ is an abbreviation for the expression (4.1).

The equations of motion follow from the variation of the action with
respect to $W_k$ and $Z^k\,$:
\begin{eqnarray}
0 & = &  \frac{\delta S^{\rm ND}}{\delta W_k} =  \! -|\gamma|^{-1/2}
{Z^k}_{,\alpha}l^{\alpha},\\
0 & = & \frac{\delta S^{\rm ND}}{\delta Z^k} = 
 \left(|\gamma|^{1/2} W_k l^{\alpha} \right)\!_{,\alpha}\,.
\end{eqnarray}
By multiplying Eq. (4.6) by $W_k \,$, we learn that $l^\alpha$ is a
null vector field:
\begin{equation}
l^{\alpha} l_{\alpha} = 0 \,.
\end{equation}
Equations (4.6) reassert that $Z^k$ are comoving coordinates.
Equation (4.7) tells us that the three currents
\begin{equation}
J^{\alpha} _{k} =  W_{k}\, l^{\alpha}
\end{equation}
satisfy the continuity equations
\begin{eqnarray}
\nabla _{\alpha}J^{\alpha}_k = \nabla _{\alpha}( W_k l^{\alpha})=0\,.
\end{eqnarray}

Because each of the three covectors ${Z^k}_{,\alpha}$ is perpendicular
to the null vector $l^\alpha$, none of them can be timelike. If only
one of the coefficients $W_k$ in the decomposition (4.1) of $l^\alpha$
does not vanish, the congruence is hypersurface orthogonal
(cf. Eq. (3.28)) and thus nontwisting. The covector ${Z^k}_{,\alpha}$
is then null. In the general case of a twisting congruence, all three
covectors ${Z^k}_{,\alpha}$ must be spacelike: If any
${Z^{k}}_{,\alpha}\,$, $\,k$ fixed, were null in an open neighborhood
${\cal U}$, then $l^{\alpha} {Z^k}_{,\alpha} = 0$ would imply
$l_{\alpha}
\propto {Z^k}_{,\alpha}$ in ${\cal U}$, and the congruence would not be
twisting in ${\cal U}$. Some covectors ${Z^k}_{,\alpha}$ can possibly
become null only in lower--dimensional $(d=0,1,2)$ regions of ${\cal
M}$. A covector ${Z^k}_{,\alpha}\,$, $\,k$ fixed, can also become null
on a 3--dimensional null hypersurface ${Z^k} = const$ on which the two
remaining coefficients $W_i\,$, $i \neq k \,$, vanish: ${W_i}
=0\,$. Then, of course, $l_{\alpha} \propto {Z^k}_{,\alpha}$
simultaneously lies in this hypersurface and is orthogonal to it.

In Appendix C, we give two examples of twisting null congruences (one
of them is the familiar ingoing principal null congruence in the Kerr
spacetime), and illustrate the decomposition (4.1) of their tangent
null covectors. The spacelike character of the covectors
${Z^k}_{,\alpha}$ is exhibited everywhere except in regions where the
twist vanishes.

It now becomes understandable why it would not be useful to represent
$l_\alpha$ by more than six potentials. If, say, we wrote $l_\alpha =
W_s {Z^s}_{,\alpha}\,$, $\,s = 1,2,3,4$, one of the spatial vectors
${Z^s}_{,\alpha}$ could always be written as a linear combination of
the remaining three vectors ${Z^k}_{,\alpha}\,$, $\,k = 1,2,3$, and the
decomposition (4.1) would be regained.

The variation of the action (4.4), (4.5) with respect to the metric
$\gamma_{\alpha \beta}$ yields the energy--momentum tensor
\begin{equation}
T^{\alpha\beta}=2|\gamma|^{-1/2}\,\delta S^{\rm ND}/\delta 
\gamma_{\alpha\beta}\,.
\end{equation}
Because $l^\alpha$ is null, Eq. (4.8), this tensor has the structure
(2.13).  The Pfaff form (4.1) satisfies the identity
\begin{equation}
\nabla_{\beta} (l_{\alpha} l^{\beta}) 
= -W_{k,\alpha}({Z^k}_{,\beta}l^\beta)
+ {Z^k}_{,\alpha} \nabla_{\beta}(W_k l^{\beta}) + \frac{1}{2} \nabla_
{\alpha} (l_{\beta} l^{\beta}) \,.
\end{equation}
The equations of motion (4.6)--(4.8) then imply the energy--momentum
conservation law. In fact, it is well known that the energy--momentum
conservation follows from the equations of motion because of the
invariance of the action (4.4), (4.5) under spacetime diffeomorphisms
(see, e.g.,\cite{B+K}). We have already seen that the energy--momentum
conservation implies that the particles of the null dust move along
geodesics, Eq. (2.14). This demonstrates that our action (4.4)--(4.5)
correctly reproduces the motion of the null dust on a given background
$({\cal M}, \gamma)\,$.

The null dust is coupled to gravity by adding its action $S^{\rm ND}$
to the Hilbert action
\begin{equation}
S^{\rm G}[\gamma_{\alpha\beta}]=\int_{{\cal M}}^{}d^4y \,  L^{\rm G} \,,
\end{equation}
\begin{equation}
L^{\rm G} =  |\gamma|^{1/2} R(y;\,\gamma]
\end{equation}
constructed from the curvature scalar $ R(y;\,\gamma]
\,$.\footnote[2]{The mixed brackets in $ R(y;\,\gamma]$
indicate that the curvature scalar $R$ is a function of $y$ and a
functional of $\gamma_{\alpha\beta}(y')\,$. This convention is used
throughout the paper.}  The variation of the total action $S = S^{\rm
G} + S^{\rm ND}$ with respect to the metric $\gamma_{\alpha \beta}$
yields the Einstein law of gravitation (2.3) with the null dust source
(4.11). The conservation law (2.4) then follows independently of the
equations of motion directly from Eqs. (2.3) through the Bianchi
identities.

\section{Special parametrizations and null dust action}
The geodesic equation (2.14) which follows from the action (4.4)--(4.5)
is not given in affine parametrization. Rather, the vector field
$l^{\alpha}(y)$ is chosen such that it absorbs the mass distribution
$M$ of the dust and leads thereby to the energy--momentum tensor
(2.13).  Let us now show that from any solution $W_{k}(y)$ and
$Z^{k}(y)$ of the Euler equations (4.6)--(4.7) of the action
(4.4)--(4.5) one can construct a vector field $k^{\alpha}(y)$ given in
generic affine parametrization.

Start on a spacelike hypersurface $\Sigma$ transverse to the dust
lines $l^{\alpha}$. Parametrize $\Sigma$ by the dust space coordinates
$z^k$ of points $z\in \cal S$. As long as there is any dust on
$\Sigma\,$, $W_{k}(z)$ cannot be a zero covector in $T^* \cal
S$. Choose an arbitrary vector field ${\bf \Lambda}^{k}(z) \in T \cal
S$ such that ${\bf \Lambda}^{k}(z)  W_{k}(z) > 0\,$. Evolve the fields
$Z^{k}(y), W_{k}(y)$ from their initial values $z^k$ and $W_{k}(z)$
on $\Sigma$ by the Euler equations (4.6)--(4.7) and define
\begin{equation}
{\bf \Lambda} (y)\, \mbox{:=}\, \left( {\bf \Lambda}^{k}
 {\bf (}Z(y){\bf )}\, W_{k}(y) \right) ^{-1} \,.
\end{equation}
The Euler equations imply that
\begin{equation}
\nabla_{\alpha} ( {\bf \Lambda}^{-1} l^{\alpha}) = 0 \,.
\end{equation}
By comparing Eq. (5.2) with Eq. (3.11), one sees that ${\bf
\Lambda}(y)$ is a scaling factor (3.10) which takes $l^{\alpha}$ into
an affinely parametrized $k^{\alpha}\,$. We already know, Eq. (3.23),
that the most general scaling factor $\Lambda (y)$ can differ from our
particular scaling factor ${\bf \Lambda}(y)$ only by a multiplicative
function $\Lambda_{0}(z)$ of comoving coordinates:
\begin{equation}
\Lambda (y) = \Lambda_{0} {\bf(} Z(y){\bf)}\, {\bf \Lambda} (y) \,.
\end{equation}

Equations (5.1), (5.3) specify an algebraic procedure by which, from
any solution $Z^{k}(y)\,$, $W_{k}(y)$ of the Euler equations
(4.6)--(4.7), one can construct the most general scaling factor
$\Lambda (y)$ which takes the covector field $l_{\alpha} = W_{k} 
{Z^k}_{,\alpha}$ into a covector field
\begin{equation}
k_{\alpha}(y) =  \Lambda (y) l_{\alpha}(y) 
\end{equation}
in affine parametrization. Equation (5.4) simultaneously tells us how
to scale the potentials $W_k$ into the corresponding potentials $w_k$
of the Pfaff form of $k_\alpha \,$:
\begin{equation}
k_\alpha = w_{k} {Z^k}_{,\alpha} \,, \;\; {\rm with} \;\; w_{k}=
\Lambda W_k \,.
\end{equation}

The mass distribution 
\begin{equation}
M={\Lambda}^{-2}
\end{equation}
associated with the affine parametrization (5.4) satisfies a
continuity equation (3.17). The potentials $W_k$ associated with
$l_{\alpha}$ also satisfy the continuity equation (4.10). However,
because in general $\nabla_{\alpha} l^{\alpha} \neq 0 \,$, the
potentials $W_k$ do not stay the same along the dust lines:
\begin{equation}
l^{\alpha} \nabla_{\alpha} W_{k} \neq 0 \,.
\end{equation}
On the other hand, by virtue of the continuity equations (4.10) {\em and}
(3.17), the potentials $w_k$ associated with an affinely parametrized
$k_\alpha$ of Eq. (5.4) do stay the same along the dust lines:
\begin{equation}
k^{\alpha} \nabla_{\alpha} w_{k} = 0 \,.
\end{equation}
Equations (3.7), (3.9) and (5.5) enable us to interpret $w_k$
geometrically as projections of the null field $k_\alpha$ into
the hypersurfaces $y^{\alpha} = \Upsilon^{\alpha}_{\rm AFF}(v, z)$,
$v=const$, of affine foliation:
\begin{equation}
w_{k} = k_{\alpha} \frac{\partial  \Upsilon^{\alpha}_{\rm AFF}(v, z)}
{\partial z^k} \,.
\end{equation}
Notice that while $k_{\alpha}$ in affine parametrization is built
from a solution $Z^{k}(y)\,$, $W_{k}(y)$ of the Euler equations by
differentiations (5.5) and algebraic manipulations (5.1), (5.3), the
construction of the affine parameter $v=V(y)$ itself requires solving
a differential equation $k^{\alpha} V_{,\alpha}=1\,$, i.e., an
integration (3.19).

The other special parameters, the luminosity distance (3.33) and the
parallax distance (3.34), can be obtained from $Z^{k}(y)\,$, $W_{k}(y)$
by algebraic operations and differentiations. The luminosity distance
$\Lambda (y)$ is simply the scaling factor (5.3). The parallax distance
$p$ is the reciprocal value of
\begin{equation}
\theta = \Lambda ( \nabla_{\alpha} l^{\alpha}) =
\Lambda |\gamma|^{-1/2} \left(|\gamma|^{1/2}\,\gamma^{\alpha \beta}
 W_k {Z^k}_{,\beta} \right){}_{,\alpha}\,.
\end{equation}

So far, we have shown how to construct the covector field $k_{\alpha}$
in affine parametrization from a solution $Z^{k}(y)\,$, $W_{k}(y)$ of
the Euler equations (4.6)--(4.7) of the action principle (4.4)--(4.5)
written in the $l_\alpha$ parametrization. Let us now show how to
enforce affine parametrization directly from an action
principle. Require one of the potentials $W_k$ in the action
(4.4)--(4.5), say $M\, \mbox{:=}\, W_3,$ to be positive, and drop the
index from the associated comoving coordinate: $Z\, \mbox{:=}\,
Z^3\,$. Introduce ${\rm w}_{A}\, \mbox{:=}\, W_{A}/W_3$ in place of
the remaining two potentials $W_A\,$, $A=1,2\,$, and write the
Lagrangian (4.5) in terms of the new variables $M$, $Z$, $Z^{A}$,
${\rm w}_A\,$:
\begin{equation}
L^{\rm ND} = - \frac{1}{2} |\gamma|^{1/2} M \gamma^{\alpha \beta}
k_{\alpha}k_{\beta}\,,
\end{equation}
with
\begin{equation}
k_{\alpha}\, \mbox{:=}\, Z_{,\alpha} + {\rm w}_{A} {Z^{A}}_{,\alpha}\,.
\end{equation}
The Pfaff form corresponding to $k_\alpha$ is now constructed only
from five potentials $Z$, ${Z^A}$, ${\rm w}_A\,$, though the action (4.4),
(5.11)--(5.12) still depends on six scalar variables, due to the
presence of $M$ in the Lagrangian (5.11).  By varying the action with
respect to $Z$ one obtains the continuity equation (3.17). By using
the other field equations, one easily derives Eq. (3.16) for affinely
parametrized geodesics.

The Lagrangian (4.5) is special because it leads to the simplified
form of the energy--momentum tensor, while the Lagrangian (5.11) is
special because it leads to an affinely parametrized $k^\alpha$. By
building an additional redundancy into the Lagrangian, one can reach
the generic form (2.1), (2.9) of the energy--momentum source. One
simply introduces the seventh scalar $M$ while keeping $U_\alpha$ as
the Pfaff form of six scalar fields $Z^k$, ${\rm W}_{k} \,$:
\footnote[3] {By comparing Eqs. (5.13)--(5.14) with Eqs. (4.2) and
(4.5), we see that $W_{k} = M^{1/2} {\rm W}_{k}\,$.}
\begin{equation}
L^{\rm ND} = - \frac{1}{2} |\gamma|^{1/2} M \gamma^{\alpha \beta}
U_{\alpha}U_{\beta}\,,
\end{equation}
with
\begin{equation}
U_{\alpha}\, \mbox{:=}\, {\rm W}_{k} {Z^k}_{,\alpha}\,.
\end{equation}
The new Lagrangian density and all equations of motion are then
invariant under the gauge transformation (2.10), (2.11),
\begin{equation}
 {\rm W}_{k} \rightarrow \overline{\rm W}_{k} = \Lambda  {\rm W}_{k}
\end{equation}
and
\begin{equation}
M \rightarrow \overline{M}=\Lambda^{-2} M \,,
\end{equation}
where $\Lambda(y)>0$ is an arbitrary scaling factor.

The canonical form of the action is the same whether one starts from
the original Lagrangian (4.5), (4.1) or the redundant Lagrangian
(5.13)--(5.14). The canonical variables recombine the redundant
potentials in such a way that the information about the split of
$l_\alpha$ into $M$ and $U_\alpha$ gets lost: From the canonical
variables one can reconstruct only $l_\alpha\,$. It is thus not worth
the effort to complicate the spacetime Lagrangian by striving to
achieve a superfluous generality. Having learned this lesson, we take
the spacetime action (4.4)--(4.5) with $l_\alpha$ given by Eq. (4.1) as
our starting point.

\section{Canonical description of null dust}
The familiar ADM algorithm for casting a covariant action into
Hamiltonian form works for the null dust in a similar way as for the
ordinary dust of massive particles \cite{B+K}.  One foliates the
spacetime ${\cal M}$ by spacelike hypersurfaces $\Sigma\,$,
\begin{eqnarray}
Y: {\hbox{$I$\kern-3.8pt $R$}} \times \Sigma \rightarrow {\cal M}
 \;\;\;\mbox{by}\;\;\; (t,x) \mapsto y=Y(t,x)\,.
\end{eqnarray}
In local coordinates $x^a,\;a = 1,2,3$ on $\Sigma$ and $y^\alpha$, 
$\alpha = 0,1,2,3$ on
${\cal M}$, the foliation is represented by
\begin{eqnarray}
(t,x^a) \mapsto y^{\alpha}=Y^{\alpha}(t,x^a)\,.
\end{eqnarray}
A transition from one leaf $\Sigma$ of the foliation to another is
described by the deformation vector $\dot{Y}^{\alpha}\, \mbox{:=}\,
\partial Y^{\alpha}/\partial t\,$. Its decomposition into the normal
$n^\alpha$ and tangential ${Y^\alpha}_{,a}$ directions to the leaves
yields the lapse function $N^{\perp}$ and the shift vector $N^a$:
\begin{equation}
\dot{Y}^{\alpha}=N^{\perp}n^{\alpha}+N^a {Y^\alpha}_{,a}\,.
\end{equation}
On each leaf, the spacetime metric $\gamma_{\alpha \beta}(y)$
induces the intrinsic metric
\begin{equation}
g_{ab}(t,x) = \gamma_{\alpha \beta}{\bf (}Y(t,x){\bf )}\,
{Y^\alpha}_{,a}(t,x){Y^\beta}_{,b}(t,x)\,.
\end{equation}
The spacetime metric is reconstructed as
\begin{equation}
\gamma^{\alpha\beta}=-n^{\alpha}n^{\beta}+g^{ab}
{Y^\alpha}_{,a} {Y^\beta}_{,b}\,,
\end{equation}
where $g^{ab}$ is the inverse of $g_{ab}\,$, and the determinants
$|\gamma|$ of $\gamma_{\alpha \beta}$ and $|g|$ of $g_{ab}$ are
related by
\begin{equation}
|\gamma|^{1/2}=N^{\perp }|g|^{1/2}.
\end{equation}

Scalar fields on ${\cal M}$, such as the null--dust variables $Z^k$,
$W_k$ can be pulled back to $ {\hbox{$I$\kern-3.8pt $R$}} \times
\Sigma$ by the mapping (6.1). By using Eq. (6.3) we obtain
\begin{equation}
{Z^k}_{,\alpha}n^{\alpha}=(N^{\perp })^{-1} V^{k}\,,
\end{equation}
where we have introduced the normal velocities
\begin{equation}
V^{k}\, \mbox{:=}\, \dot{Z}^k - {Z^k}_{,a} N^a ,\;\;\;
{Z^k}_{,a}={Z^k}_{,\alpha}{Y^\alpha}_{,a}\,.
\end{equation}
This allows us to write the null--dust action (4.4)--(4.5), with
$l_\alpha$ given by (4.1), as an integral over ${\hbox{$I$\kern-3.8pt
$R$}}\times\Sigma$, i.e., in the (3+1)--split form:
\begin{equation}
S^{\rm
ND}[Z^k,W_k\,;\;g_{ab},N^{\perp},N^a]=\int_{{\hbox{$\scriptstyle
I$\kern-2.4pt $\scriptstyle R$}}} dt \int_\Sigma d^3x \; L^{\rm ND}.
\end{equation}

The Lagrangian density $L^{\rm ND}$ on $ {\hbox{$I$\kern-3.8pt $R$}}
\times \Sigma $ is a quadratic form of the Lagrange multipliers
$W_k\,$:
\begin{equation}
L^{\rm ND} = \frac{1}{2} |g|^{1/2} \left( (N^{\perp})^{-1}
V^{i} V^{j} - N^{\perp} g^{ij} \right) W_{i} W_{j} \,.
\end{equation}
The metric
\begin{equation}
 g^{ij}(t,x) = g^{ab} {Z^i}_{,a} {Z^j}_{,b}
\end{equation}
is the induced metric on $\Sigma$ expressed in the basis ${Z^i}_{,a}$
of comoving coordinates ${Z^i}$.

By varying the action with respect to $W_i\,$, we get a system of
linear homogeneous equations for $W_j\,$:
\begin{equation}
\left( g^{ij}  - (N^{\perp})^{-2} V^{i} V^{j} \right)\,
W_{j} = 0 \,.
\end{equation}
This has a nontrivial solution only if the determinant
\begin{equation}
\det \left( g^{ij}  - (N^{\perp})^{-2} V^{i} V^{j} \right) =
\left( 1 - (N^{\perp})^{-2} g_{ij}V^{i} V^{j} \right)\, \det 
{\bf (} g^{ij} {\bf )}
\end{equation}
vanishes. This imposes the constraint
\begin{equation}
g_{ij}V^{i} V^{j} - (N^{\perp})^{2} = 0
\end{equation}
on the velocities $\dot{Z}^i$. (Here, $g_{ij}$ is the inverse of
$g^{ij}$. We can use it for lowering the dust space indices.)

If the constraint (6.14) is satisfied, Eq. (6.12) has a solution $W_{j}
\propto V_{j}\,$. Of course, the homogeneous equation (6.12) 
determines only the direction of $W_j \,$, leaving 
$W= g^{ij}W_{i}W_{j}$ undetermined. We write the general
solution in the form
\begin{equation}
W_{j}=\sqrt{W}\, V_{j}\,(V_{i}V^{i})^{-1/2}\,,
\end{equation}
where $W$ is an arbitrary positive factor.

By substituting this solution (6.15) back into the Lagrangian (6.10),
we eliminate from the action the multipliers $W_j\,$, replacing them
by a single multiplier $W\,$:
\begin{equation}
L^{\rm ND} = \frac{1}{2} |g|^{1/2} \,W\left( (N^{\perp})^{-1} g_{ij}
V^{i} V^{j} - N^{\perp} \right). 
\end{equation}
 The reduced action is a functional of $W$ and $Z^k$. Its variation
with respect to $W$ reproduces the constraint (6.14) which enabled us
to express $W_j$ in terms of $W$ and $Z^j$, $\dot{Z}^{j}$,
Eq. (6.15). Its variation with respect to $Z^k$ gives an equation
which, modulo the constraint (6.14) and Eq. (6.15) considered as a
definition of $W_j \,,\,$ is equivalent to the equations of motion
obtained by varying the original action (6.9)--(6.10) with respect to
$Z^k$. The reduced action
\begin{equation}
S^{\rm ND}[ Z^{k}, W;\, g_{ab}, N^{\perp}, N^{a} ] = 
\int_{{\hbox{$\scriptstyle
I$\kern-2.4pt $\scriptstyle R$}}} dt \int_\Sigma d^3x \, L^{\rm ND}
\end{equation}
with the Lagrangian (6.16) is thus entirely equivalent to the original
action (6.9) with the Lagrangian (6.10).

In order to bring the reduced action to canonical form, we perform the
Legendre dual transformation from $(Z^{k}, \dot{Z}^{k})$ to $(Z^{k},
P_{k})$, leaving $W$ as a multiplier. First, we introduce the momenta
\begin{equation}
P_{k} \,\mbox {:=}\, \frac{\partial L^{\rm ND}}{\partial \dot{Z}^{k}}
= |g|^{1/2}W(N^{\perp})^{-1} V_{k}\,.
\end{equation}

To clarify their physical meaning, we return to the definition (6.15)
of $W_{j}\,$, the decomposition (4.1) of $l_{\alpha}\,$, and
Eqs. (6.7)--(6.8) for the normal velocity. In this way we learn that
$P_k$ are normal projections of the currents $J^{\alpha}_{k}$
introduced in Eq. (4.9):
\begin{equation}
P_{k} = |g|^{1/2} J^{\alpha}_{k} n_{\alpha}\,.
\end{equation}

Equation (6.18) can be inverted to obtain the velocities
\begin{equation}
\dot{Z}^{k} = N^{\perp} |g|^{-1/2} W^{-1} g^{kj} P_{j} +
N^{a} {Z^k}_{,a}\,.
\end{equation}
This leads to the Hamiltonian
\begin{equation}
H^{\rm ND} \,\mbox{:=}\, P_{k}\dot{Z}^{k} - L^{\rm ND} = N^{\perp}
H^{\rm ND}_{\perp} + N^{a} H^{\rm ND}_{a}
\end{equation}
which is a linear combination of the momentum density
\begin{equation}
 H^{\rm ND}_{a} = P_{k}{Z^k}_{,a}
\end{equation}
and the energy density
\begin{eqnarray}
H^{\rm ND}_{\perp} & =& \frac{1}{2} W^{-1} |g|^{-1/2} g^{ij}P_{i}P_{j} +
\frac{1}{2} W |g|^{1/2} \\
{} & =& \frac{1}{2} W^{-1} |g|^{-1/2} g^{ab}H^{\rm ND}_{a}
H_{b}^{\rm ND} +\frac{1}{2} W |g|^{1/2}
\end{eqnarray}
of the dust. The canonical form of the action then reads
\begin{equation}
S^{\rm ND}[ Z^{k}, P_{k}, W; g_{ab}, N^{\perp}, N^{a}] =
\int_{{\hbox{$\scriptstyle
I$\kern-2.4pt $\scriptstyle R$}}} dt \int_\Sigma d^3x \,
\left( P_{k}\dot{Z}^{k} - N^{\perp} H_{\perp}^{\rm ND}
 - N^{a} H_{a}^{\rm ND} \right),
\end{equation}
where
$H_{a}^{\rm ND}$ and $H_{\perp}^{\rm ND}$ are given by Eqs. (6.22)
and (6.24).

At this stage, we are finally able to eliminate the last remaining
multiplier $W$. By varying the action (6.22)--(6.25) with respect to
$W$, we obtain an equation
\begin{equation}
\frac{\delta S^{\rm ND}}{\delta W} = - N^{\perp} \frac{\partial
H_{\perp}^{\rm ND}}{\partial W} = 0
\end{equation}
which determines $W$ in terms of the canonical data:
\begin{equation}
W =|g|^{-1/2}\,\sqrt{g^{ij}P_{i}P_{j}}=|g|^{-1/2}\,
\sqrt{g^{ab}H^{\rm ND}_{a}H^{\rm ND}_{b}}\,.
\end{equation}
By substituting this solution back into $H_{\perp}^{\rm
ND}$ we obtain
\begin{equation}
H_{\perp}^{\rm ND} = \sqrt{g^{ab}H^{\rm ND}_{a}H^{\rm ND}_{b}}\,.  
\end{equation}
We see that $W$ is just the scalar form $W= |g|^{-1/2}H_{\perp}^{\rm
ND}$ of the Hamiltonian density $H_{\perp}^{\rm ND}$. The final
expressions (6.22) for the momentum density and (6.28) for the energy
density are simple: The form of the momentum density is dictated by
the requirement that it generate the Lie derivative change of the
scalars $Z^{k}(x)$ and scalar densities $P_{k}(x)$ under spatial
diffeomorphisms LDiff$\Sigma\,$\cite{kk-kh}. The energy density is the
norm of the momentum density with respect to the spatial metric. The
resulting reduced canonical action $S^{\rm ND}[ Z^{k}, P_{k};\,
g_{ab}, N^{\perp}, N^{a}]$, with (6.22), (6.28) yields the Hamilton
equations for $Z^{k}(t,x)$ and $P_{k}(t,x)$. These describe the
evolution of the null dust on a given geometrical background
$\gamma_{\alpha \beta}
\leftrightarrow \left( N^{\perp}(t,x),\;  N^{a}(t,x),\; g_{ab}(t,x)\,
\right)$. \footnote[4]{Our Lagrangian and Hamiltonian formalism can
easily be generalized to several mutually noninteracting species
(streams) of null dust. This may be useful for the canonical treatment
of spherical collapse, in which the ingoing null dust is turned into
an outgoing null dust at the center of symmetry, or for the canonical
treatment of models involving colliding streams of dust with plane or
cylindrical symmetry (see references in \cite{col}).}

From the solution of the Hamilton equations we can reconstruct the
null vector $l^\alpha$ which provides the spacetime description of
the dust. It holds that
\begin{equation}
l^{\alpha} = l^{\perp} n^{\alpha} + l^{a} {Y^\alpha}_{,a}\,,
\end{equation}
where $ l^{\perp}$ and $ l_{a}$ are expressed as functions of the
canonical variables:
\begin{equation}
l^{\perp} = -l_{\alpha} n^{\alpha} = -W^{1/2},
\end{equation}
\begin{equation}
l_{a} = l_{\alpha} {Y^\alpha}_{,a} = W^{-1/2} |g|^{-1/2}
H^{\rm ND}_{a}\,.
\end{equation}
Here, of course, $W$ stands for the scalar form (6.27) of the
energy--momentum density. One can check that $l^{\alpha}$ is a
null vector by virtue of its construction (6.29)--(6.31):
\begin{equation}
l^{\alpha} l_{\alpha} =-( l^{\perp})^{2} + l^{a} l_{a} =0\,.
\end{equation}

The background variables $ N^{\perp}(t,x)$, $N^{a}(t,x)$ and
$g_{ab}(t,x)$ in the dust action $S^{\rm ND}[ Z^{k}, P_{k}\,;\;
g_{ab}\,, N^{\perp}, N^{a}]$ are not to be varied. The Hamiltonian
formalism for null dust on a given background is thus entirely
unconstrained. To couple null dust to geometry, we must add its action
$S^{\rm ND}$ to the gravitational Dirac--ADM action
\begin{equation}
S^{\rm G}\left[g_{ab}, p^{ab} ; N^{\perp}, N^{a}\right]=
\int_{{\hbox{$\scriptstyle
I$\kern-2.4pt $\scriptstyle R$}}} dt \int_\Sigma d^3x \,
\left( p^{ab}\dot{g}_{ab} - N^{\perp} H_{\perp}^{\rm G}
 - N^{a} H_{a}^{\rm G} \right)
\end{equation}
with the standard gravitational super--Hamiltonian and supermomentum
densities
\begin{eqnarray}  
H^{\rm G}_\perp(x;\,g_{ab},p^{ab}] & = & G_{abcd}(x;g)\; 
p^{ab}(x) p^{cd}(x) - |g|^{1/2} R(x;\,g] \ , \\ G_{abcd} & = &
\frac{1}{2} |g|^{-1/2} (g_{ac} g_{bd} + g_{ad}g_{bc} -
g_{ab} g_{cd} ) \ , \\ H^{\rm G}_a(x;\,g_{ab},p^{ab}] & = &
-2D_b p^b_a(x) \, ; \end{eqnarray}
here, $D_b$ is the spatial covariant derivative. 

The variation of the total action with respect to the lapse
$N^{\perp}$ and the shift $N^{a}$ then leads to the familiar
Hamiltonian and momentum constraints
\begin{eqnarray}  H_\perp & := & H^{\rm G}_\perp +
     H^{\rm ND}_\perp = 0 \ ,   \\ 
    H_a & := & H_a^{\rm G} + H^{\rm ND}_a = 0 
    \end{eqnarray}
for the coupled system. 

\section{Null dust constraints that generate a Lie algebra}
By using the supermomentum constraint, one can replace the momentum
density $H^{\rm ND}_{a}$ of the dust by the gravitational density
$H^{\rm G}_{a}$ in the expression (6.28) for the dust energy density
$H^{\rm ND}_{\perp }\,$. This brings the constraint system (6.38),
(6.37) into an equivalent form (6.38) and
\begin{equation}
H_{\perp} \,\mbox{:=}\, H^{\rm G}_{\perp} + \sqrt{g^{ab}H^{\rm G}_{a}
H^{\rm G}_{b}} = 0\,.
\end{equation}
Only the supermomentum constraint (6.38) contains the dust variables.
The new Hamiltonian constraint (7.1) is constructed solely from the
gravitational variables $g_{ab}\,,\,p^{ab}$. Alternatively, one can
get rid of an inconvenient square root by rewriting Eq. (7.1) in the
form
\begin{equation}
G \,\mbox{:=}\,(H^{\rm G}_{\perp})^{2} - g^{ab}H^{\rm G}_{a}
H^{\rm G}_{b} = 0\,.
\end{equation}
Under the positivity condition
\begin{equation}
-H^{\rm ND}_{\perp} = H^{\rm G}_{\perp} \leq 0\,,
\end{equation}
the constraint (7.2) is equivalent to the constraint (7.1).

Brown and Kucha\v{r} \cite{B+K} proved a remarkable fact that the 
densities (7.2) have strongly vanishing Poisson brackets:
\begin{equation}
\{ G(x),\,G(x')\}=0\,.
\end{equation}
By coupling gravity to other simple sources, Kucha\v{r} and Romano
\cite{K+R} and Brown and Marolf \cite{B+M} produced other densitized
expressions constructed from the scalar variables $g^{-1/2}H^{\rm G}
_\perp$ and $g^{-1}g^{ab}H^{\rm G}_{a}H^{\rm G}_{b}$ which also have
strongly vanishing Poisson brackets. Markopoulou \cite{M} posed the
question what is the most general density
\begin{equation}
F=g^{w/2} {\rm F}(g^{-1/2}H^{\rm G}_{\perp}\,,\; g^{-1}g^{ab}H^{\rm
G}_{a}H^{\rm G}_{b})
\end{equation}
of weight $w$ constructed from these variables which has the strongly
vanishing Poisson \mbox{brackets}
\begin{equation}
\{ F(x),\,F(x') \}=0\,.
\end{equation}
She found an algorithm for generating all such densities. The density
(7.2) of weight 2 still seems to be the simplest. Among others, there
is the scalar form
\begin{equation}
G_{\sqrt{}}\; \mbox{:=}\, g^{-1/2} \left( H^{\rm G}_{\perp} +
\sqrt{g^{ab} H^{\rm G}_{a}H^{\rm G}_{b}}\, \right) = 0 
\end{equation}
of the constraint (7.1) which, as we have just seen, describes null dust.

The constraints $H^{\rm G}_{a}=0=H^{\rm G}_{\perp}$ of vacuum gravity
can be replaced by an alternative system
\begin{equation}
H^{\rm G}_{a}=0=G\;\;\;\;\left(\,{\rm or}\;\;H^{\rm
G}_{a}=0=G_{\sqrt{}}\,\right)\,.
\end{equation}
Unlike the original constraints, $H^{\rm G}_{a}$ and $G$ (or $H^{\rm
G}_{a}$ and $G_{\sqrt{}}\,$) generate a
true Lie algebra. Unfortunately, in vacuum gravity the new constraints
(7.2) (and similarly (7.7)) do not generate the evolution of the
geometric data $g_{ab}\,,\; p^{ab}$ into a Ricci--flat
spacetime. Expression (7.2) is flawed because its Hamiltonian vector
field vanishes on the constraint surface (7.8), while the expression
(7.7) is flawed because its Hamiltonian vector field is ill--defined
for $H^{\rm G}_{a}=0\,$.
 
No such difficulty exists for null dust. The momentum constraint
(6.38) is different from the vacuum constraint $H^{\rm G}_{a}=0$ and,
as long as there is any dust at the point in question, $H^{\rm
ND}_{a}$ and hence $H^{\rm G}_{a}$ cannot vanish. The Hamiltonian
vector fields of the dynamical variables (7.2) or (7.7) then do not
vanish on the constraint surface (6.37)--(6.38) of the null dust
coupled to geometry. The new constraints (7.2) or (7.7) correctly
generate the evolution of geometry produced by null dust. Moreover, as
in vacuum spacetime, the constraints (6.38) and (7.2), or (6.38) and
(7.7), generate a true Lie algebra. It is thus advantageous to bring
the constraints to one of these forms before attempting to quantize
the coupled system.

Why is it that the presence of the null dust does not affect
Eqs. (7.2) and (7.7) that hold in vacuum gravity? The energy--momentum
tensor of the null dust satisfies the condition
\begin{equation}
{T^\alpha}_{\gamma} T^{\gamma \beta} =0\,.
\end{equation}
Conversely, any symmetric tensor $T^{\alpha \beta}$ which satisfies
Eq. (7.9) must either vanish, or there exists a null vector $l^\alpha$
such that
\begin{equation}
T^{\alpha \beta}= l^{\alpha} l^{\beta}.
\end{equation}
The Einstein law of gravitation (2.3) then implies that $l^\alpha$ 
is a geodesic vector field, i.e., the Euler equations of motion for
the null dust. The simple tensor equation
\begin{equation}
{G^\alpha}_{\gamma} G^{\gamma \beta}=0
\end{equation}
imposed on the Einstein tensor thus ensures that the geometry $\gamma
_{\alpha \beta}$ is necessarily produced by null dust according to
Einstein's law of gravitation.\footnote[5] {Equation (7.11) is perhaps
the simplest example of the Rainich--type geometrization of a source
field. The general task is to find equations for the Einstein tensor
which are equivalent to the Einstein law of gravitation together with
the field equations for a given source. The problem was first
formulated for the Einstein--Maxwell system by Rainich and solved by
him under the assumption that the electromagnetic field is not
algebraically special (null) \cite{rainich}, \cite{M+W}. The Rainich
problem for the null electromagnetic field was solved by Hlavat\'{y}
\cite{hlav}. The much simpler scalar field case was analyzed by Peres
\cite{per} and by Kucha\v{r} \cite{kk-rain}. The spinor field 
was treated by Kucha\v{r} \cite{kk-neu} and the Proca field by
Bi\v{c}\'{a}k \cite{bica}. }

The $\perp \perp$ projection of Eq. (7.11) gives
\begin{equation}
(G_{\perp \perp})^{2} - g^{ab}G_{\perp a}G_{\perp b}=0\,.
\end{equation}
Because the $\perp \perp$ and $\perp \parallel $ projections of the
Einstein tensor yield the gravitational super--Hamiltonian and
supermomentum \cite{kk-kh},
\begin{equation}
G_{\perp \perp}= - \frac{1}{2} g^{-1/2} H^{\rm G}_{\perp}\,,\;\;\;
G_{\perp a}= \frac{1}{2} g^{-1/2} H^{\rm G}_{a}\,,
\end{equation}
Eq. (7.12) is equivalent to the constraint (7.2). We have already
noticed that under the energy positivity condition (7.3) the constraint
(7.2) is equivalent to the constraint (7.7). The Rainich--type
condition (7.11) thus connects the new form (7.2) or (7.7) of
the Hamiltonian constraint with the spacetime picture.

\section{Constraint quantization of geometry coupled to null dust}
We have cast the constraint system for geometry coupled to null dust
into a form in which it generates a Lie algebra. In this process, the
Hamiltonian constraint has been replaced either by the constraint
(7.2) or by the constraint (7.7). Either of these constraints have
vanishing Poisson brackets (7.4). The momentum constraint is left in
its original form (6.38) and (6.22):
\begin{equation}
H_{a}(x) \,\mbox{:=}\, P_{k}(x){Z^k}_{,a}(x) + H^{\rm G}_{a}(x)=0\,.
\end{equation}
The momentum constraints (8.1) close in the way characteristic for the
Lie algebra LDiff$ \Sigma$ of the diffeomorphism group Diff$
\Sigma$. The Poisson brackets of $G(x)$ ${\bf (}\, {\rm or}\;
G_{\sqrt{}}\,(x)\,{\bf )}$ with $H_{a}(x')$ close into $G(x)$ ${\bf (}
\, {\rm or}\;G_{\sqrt{}}\,(x)\,{\bf )}$ in the way which reflects
the transformation behavior of $G(x)$ ${\bf (} \,{\rm
or}\;G_{\sqrt{}}\,(x)\,{\bf )}\,$ under spatial diffeomorphisms Diff$
\Sigma\,$: $G(x)$ is a density of weight 2, while $G_{\sqrt{}}\,(x)$
is a scalar.

As for ordinary dust, the constraint system can be vastly simplified
by the introduction of an alternative set of canonical variables which
reflect the fact that the dust particles define a preferred system
of coordinates on $\Sigma$. The mapping $Z:\Sigma\to{\cal S}$ which,
in local coordinates, assumes the form
\begin{equation}
z^{k}=Z^{k}(x^{a})\,,
\end{equation}
takes the tensorial variables $g_{ab}(x)$ and $p^{ab}(x)$ on $\Sigma$
into corresponding tensors ${\bbox{g}}_{ij}(z)$ and
${\bbox{p}}^{ij}(z)$ on the dust space ${\cal S}\,$:
\begin{eqnarray} 
   {\bbox{g}}_{ij}(z) & := & {X^a}_{,i}(z) {X^b}_{,j}(z) 
       \, g_{ab}{\bf (}X(z){\bf )} \ , \\
    {\bbox{p}}^{ij}(z) & := & \left|\frac{\partial X(z)} 
     {\partial z}\right| {Z^i}_{,a}{\bf (}X(z){\bf )}
      {Z^j}_{,b}{\bf (}X(z){\bf )} \, p^{ab}{\bf (}X(z){\bf )} \ .
\end{eqnarray} 
Here, the $t$--dependent mapping $\, X:{\cal S}\to\Sigma\,$ is simply
the inverse of $Z$,
\begin{equation}
X := Z^{-1} , 
\end{equation} 
and $|\partial X(z) / \partial z|$ is the Jacobian for the 
change of variables $x^a = X^a(z)$.

We rewrite the supermomentum constraint (8.1) in the form
\begin{equation}
H_{\uparrow k}(x) \,\mbox{:=}\, H_a(x)Z^a_k (x) =
P_k(x) +  H^{\rm G}_a(x)Z^a_k (x) = 0\,.
\end{equation}
Here,
\begin{equation}
Z^a_k (x) \,\mbox{:=}\, {X^a}_{,k}{\bf (}Z(x){\bf )}
\end{equation}
is the inverse matrix to ${Z^k}_{,a}(x)\,$:
\begin{equation}
{Z^j}_{,a}(x)Z^a_k(x)=\delta^j_k \,.
\end{equation}
The new supermomentum $H_{\uparrow k}(x)$ smeared by a new shift
$N^{\uparrow k}(x)$,
\begin{equation}
H_{\uparrow }[{\vec N}^{\uparrow}] \,\mbox{:=}\, 
\int_{\Sigma} d^{3}(x)\, N^{\uparrow k}(x)H_{\uparrow k}(x)\,,
\end{equation}
generates through the Poisson bracket the change
\begin{equation}
\dot{Z}^k (x) \,\mbox{:=}\, \left\{ Z^k (x),\,H_{\uparrow}[
{\vec N}^{\uparrow}] \right\} = N^{\uparrow k}(x)
\end{equation}
of the dust coordinates $Z^k (x)$ by the amount $ N^{\uparrow
k}(x)\,$\cite{B+K}.

One can prove that the ${\cal S}$--variables $ {\bbox{g}}_{ij}(z)$, $
{\bbox{p}}^{ij}(z)$ along with the dust frame variables $Z^k(x)$ and
the new supermomentum $H_{\uparrow k}(x)$ form a canonical chart
\cite{B+K}. In particular, this means that the new constraint
functions ${\rm P}_k(x)\,\mbox{:=}\,H_{\uparrow k}(x)$ have vanishing
Poisson brackets among themselves and are the momenta ${\rm P}_k(x)$
canonically conjugate to the dust frame variables $Z^k(x)$. Further,
because the Poisson brackets of the $\cal S$--tensors $
{\bbox{g}}_{ij}(z)$, $ {\bbox{p}}^{ij}(z)$ with the smeared
supermomentum (8.9) vanish, these $\cal S$--tensors are invariant
under the shifts (8.6).

In terms of the new canonical variables $ {\bbox{g}}_{ij}(z)$, $
{\bbox{p}}^{ij}(z)$ and $Z^k(x)$, ${\rm P}_k(x)$ the momentum
constraint (8.6) reduces to the condition that the canonical momentum
${\rm P}_k(x)$ vanishes:
\begin{equation}
{\rm P}_k(x)=0\,.
\end{equation}
The Hamiltonian constraints (7.2) or (7.7) can then be mapped to the
dust space $\cal S$ according to their weight:
\begin{eqnarray}
{\bf G}(z)& \,\mbox{:=}\,& \left|\frac{\partial X(z)} 
     {\partial z}\right|^{2} G{\bf (}X(z){\bf )} =0\,,\\
{\bf G}_{\sqrt{}}(z)& \,\mbox{:=}\,& G_{\sqrt{}}{\bf (}X(z){\bf )} =0\,.
\end{eqnarray}

The $\cal S$--constraints (8.12)--(8.13) are the same functionals of
the $\cal S$--tensors $ {\bbox{g}}_{ij}(z)$ and $
{\bbox{p}}^{ij}(z)$ as the $\Sigma$--constraints (7.2) or (7.7) were
of the $\Sigma$--tensors $g_{ab}(x)$ and $p^{ab}(x)\,$. In other words,
${\bf G}(z)$ is obtained from $G(x)$ and ${\bf G}_{\sqrt{}}(z)$ is
obtained from $G_{\sqrt{}}(x)$ by replacing the $\Sigma$--tensors
$g_{ab}(x)$, $p^{ab}(x)$ by the corresponding $\cal S$--tensors $
{\bbox{g}}_{ij}(z)$, $ {\bbox{p}}^{ij}(z)$. The $\cal
S$--constraints (8.12) and (8.13) have strongly vanishing Poisson
brackets, i.e., they generate an Abelian algebra.

In the Dirac method of quantization, constraints are turned into
operators and imposed as restrictions on the state functionals of the
system. We choose to work with the dust space variables, so the
quantum states of the system are functionals ${\bbox
{\Psi}}[Z,\,{\bbox {g}}]$ of the canonical coordinates $Z^k(x)$ and
${\bbox{g}}_{ij}(z)$, and the constraint operators $\widehat{\rm P}
_{k}(x)$ and $\widehat{\bbox{G}}(z)$ (or
$\widehat{\bbox{G}}_{\sqrt{}}\,(z)\,$) are obtaned by quantizing the
classical expressions (8.11)--(8.13). The transition is easy for the
momenta (8.11) which are simply replaced by the variational derivatives
\begin{equation}
\widehat{\rm P}_{k}(x)=-i\frac{\delta}{\delta Z^{k}(x)}\;.
\end{equation}
The operators (8.14) automatically commute,
\begin{equation}
\left[ \widehat{\rm P}_{i}(x),\, \widehat{\rm P}_{j}(x')\right]=0\,.
\end{equation}
It is far from clear how to replace the remaining classical
constraints by operators which not only commute with
$\widehat{\rm P}_{k}(x)$, but also among themselves. We shall proceed
under the assumption that there exists a factor ordering and
regularization of $\widehat{\bbox{G}} \,\mbox{:=}\,{\bbox{G}}(z;\,
\widehat{\bbox{g}}_{ij}(z),\, \widehat{\bbox{p}}^{ij}(z)]$ and/or
$\widehat{\bbox{G}}_{\sqrt{}} \,\mbox{:=}\,{\bbox{G}}_{\sqrt{}}(z;\,
\widehat{\bbox{g}}_{ij}(z),\, \widehat{\bbox{p}}^{ij}(z)]$ which achieves 
this goal. If so, the constraint operators can consistently annihilate
the physical states. The momentum constraint
\begin{equation}
\widehat{\rm P}_{k}(x)\,{\bbox
{\Psi}}[Z,\,{\bbox {g}}] = 0\,,
\end{equation}
where $\widehat{\rm P}_{k}(x)$ is interpreted as the variational
derivative (8.14), means that the state functional ${\bbox
{\Psi}}[Z,\,{\bbox{g}}]$ cannot depend on $Z^{k}(x)$:
\begin{equation}
{\bbox{\Psi}}={\bbox{\Psi}}[{\bbox{g}}]\,.
\end{equation}
The constraint system is thereby reduced to a single $\infty^{3}$
nontrivial condition that $\widehat{\bbox{G}}
\,\mbox{:=}\,{\bbox{G}}(z;\, \widehat{\bbox{g}}_{ij}(z),\,
\widehat{\bbox{p}}^{ij}(z)]$, 
$ (\, {\rm or}\;\widehat{\bbox{G}}_{\sqrt{}}
\,\mbox{:=}\,{\bbox{G}}_{\sqrt{}}(z;\, \widehat{\bbox{g}}_{ij}(z),\,
\widehat{\bbox{p}}^{ij}(z)]\,)$ annihilates the state functional:
\begin{equation}
{\bbox{G}}(z;\,\widehat
{\bbox{g}}_{ij}(z),\,\widehat{\bbox{p}}^{ij}(z)]\,{\bbox
{\Psi}}[{\bbox {g}}] = 0\,.
\end{equation}
The caveats which need to be born in mind when implementing such a
formal procedure for gravity coupled to ordinary dust are carefully
spelled out in \cite{B+K}. An additional difficulty with null dust is
that there is no natural variable which would play the role of
internal time. As a result, unlike for ordinary dust, the quantum
constraint (8.18) does not have the form of a functional
Schr\"{o}dinger equation. It is thus unclear how, even formally, to
turn the space of its solutions into a Hilbert space.
\section{Comparing null dust with ordinary dust}
Ordinary dust coupled to gravity was turned into a Hamiltonian system
and formally quantized by Brown and Kucha\v{r} \cite{B+K}. This scheme
turns out to be both similar to and characteristically different from
the description of null dust given in this paper. We shall outline the
basic similarities and emphasize the differences.

The spacetime action
\begin{equation} 
S^{\rm D}[T,Z^k;\,M,W_k\,;\,\gamma_{{\alpha\beta}}]=\int_{\cal
M}d^4y\, 
L^{\rm D}(y)  
\end{equation}
of ordinary dust is constructed from eight scalar fields $Z^k$, $W_k$
and $T$, $M$. The Lagrangian density $L^{\rm D}(y)$ has the form
\begin{equation} 
 L^{\rm D} = -\frac{1}{2} |\gamma|^{1/2} M \left(
\gamma^{{\alpha\beta}} 
U_\alpha U_\beta + 1\right)  . 
\end{equation}
The four--velocity $U_{\alpha}$ is expressed as the Pfaff
form
\begin{equation} 
U_\alpha = -T_{,\alpha} + W_k {Z^k}_{,\alpha} 
\end{equation} 
of seven scalar fields $W_k$, $Z^k$ and $T$.  The matter equations of
motion are obtained by varying the dust action (9.1)--(9.3) with
respect to the state variables $M$, $W_{k}$, $T\,$ and $Z^k$:
\begin{eqnarray}  
0 & = & \frac{\delta S^{\rm D}}{\delta M} = -\frac{1}{2} 
          |\gamma|^{1/2} \left( \gamma^{{\alpha\beta}} U_\alpha
U_\beta + 1\right), 
          \\ 
 0 & = & \frac{\delta S^{\rm D}}{\delta W_k} = - 
 |\gamma|^{1/2} M {Z^k}_{,\alpha} U^\alpha  ,\\ 
    0 & = & \frac{\delta S^{\rm D}}{\delta T} = - \left( |\gamma|^{1/2} 
          M U^\alpha \right){} _{,\alpha}\,  ,\\ 
    0 & = & \frac{\delta S^{\rm D}}{\delta Z^k} = {\left( |\gamma|^{1/2} 
          M W_k U^\alpha \right)}{} _{,\alpha} \, .
\end{eqnarray} 
They lead to the interpretation of the state variables. Equation (9.5)
is analogous to Eq. (4.6) for null dust. It ensures that the three
vector fields $Z^k$ are constant along the flow lines of $U^\alpha$
and therefore their values $z^k$ can be interpreted as comoving
coordinates for the dust. Equation (9.4) ensures that the
four--velocity $U^\alpha$ is a unit timelike vector field. It is
analogous to Eq. (4.8) which guarantees that the four--velocity
$l^\alpha$ of null dust is lightlike. Equation (9.6) allows us to
interpret $M$ as the rest mass density of the dust and expresses the
law of mass conservation. It is analogous to Eq. (3.17) for the null
dust in affine parametrization. Equation (9.7) can be interpreted as
the momentum conservation law. It is analogous to Eq. (4.10) for the
null dust written again in affine parametrization. By multiplying
Eq. (9.3) by $U^\alpha$ and using the field equations (9.4)--(9.5), we
learn that
\begin{equation}
T_{,\alpha} U^{\alpha}=1\,,
\end{equation}
i.e., that $T$ is the proper time between a fiducial hypersurface
$T=0$ and an arbitrary hypersurface $T=const$ along the dust
worldlines. From Eq. (9.3) we see that the $W_k$ variables are the
projections of the four--velocity $U_\alpha$ to the hypersurfaces of
constant $T$ expressed in the dust space cobasis
${Z^k}_{,\alpha}\,$. Due to the conservation laws (9.6)--(9.7), these
projections remain the same along a flow line of $U^\alpha$. In
comparison, $W_k$ for the null dust is the component of the null
covector $l^{\alpha}$ in the dust space cobasis
${Z^k}_{,\alpha}\,$. These components are {\em not} conserved along
the flow lines, Eq. (5.7). However, when one rescales $l^{\alpha}$
into an affinely parametrized $k^\alpha$ by Eqs. (5.4)--(5.5) and
projects $k_\alpha$ into hypersurfaces of constant affine parameter
$v$, one obtains the components $w_k$ of Eq. (5.9) which {\em are}
conserved along the flow lines, Eq. (5.8).

The main difference between the actions $S^{\rm D}$ and $S^{\rm ND}$
is that the dust action depends on eight variables $T$, $M$ and $Z^k$,
$W_k\,$, while the null dust action depends only on six variables
$Z^k$, $W_k\,$. The interpretation of the variables $Z^k$ as the
comoving coordinates and $W_k$ as the projections of the
four--velocities $U^\alpha$ (or $\l^\alpha$) into hypersurfaces of
constant $T$ (or $U$) is analogous. The variables $T$ and $M$ do not
appear in the null dust action (4.1), (4.4)--(4.5). This reflects the
fact that the mass function $M$ of the null dust is not uniquely
determined and it was absorbed into the definition of $\l^\alpha$.
Similarly, the affine parameter along the null geodesics is not
uniquely determined. If one chooses to enforce the affine
parametrization by taking the null dust Lagrangian in the form (5.11),
the corresponding $M$ occurs in the action, but the Pfaff form of an
affinely parametrized $k_\alpha\,$, Eq. (5.12), contains only two
independent scalars ${\rm w}_A\,$. One can work in a totally arbitrary
parametrization by letting the Lagrangian density to depend on seven
variables $M$, $Z^k$, ${\rm W}_k$ instead of six, Eqs. (5.13)--(5.14),
but then the action becomes gauge invariant under the scalings
(5.15)--(5.16), which makes it effectively dependent only on six of
these variables.

These similarities and differences are reflected in the canonical form
of the action. For ordinary dust, the energy density $H^{\rm D}_\perp$
and momentum density $H^{\rm D}_a$ depend on {\em four} pairs of
canonical variables, $T$ and $P$, and $Z^k$, $P_k\,$. They take the
form
\begin{equation}
 H^{\rm D}_a  = PT_{,a} + P_k {Z^k}_{,a} 
\end{equation}
and
\begin{equation}
H^{\rm D}_\perp = \sqrt{P^2 + g^{ab}H_a^{\rm D} H_b^{\rm D}} \,.
\end{equation}
On the other hand, similar expressions for null dust, Eqs. (6.22) and
(6.28), depend only on {\em three} pairs of canonical variables, $Z^k$
and $P_k\,$. This difference is vital. While ordinary dust has four
degrees of freedom per space point $x\in\Sigma$, null dust has only
three.

The rest mass density $M$ of ordinary dust is directly related to the
momentum $P\,$:
\begin{equation}
M = |g|^{-1/2}\, \frac{P^2}{\sqrt{P^2 + g^{ab}H_a^{\rm D}
H_b^{\rm D}}} \,.
\end{equation} 
The mass function and affine parametrization of null dust are
ambiguous and their only invariant combination is the null vector
$l^\alpha$. This can be reconstructed from the canonical data,
Eqs. (6.29)--(6.31), rather than the mass function and the
four--velocity separately.

Formally, the momentum and energy densities (6.22) and (6.28) of the
null dust are obtained from the corresponding expressions
(9.9)--(9.10) for ordinary dust simply by putting $P=0$ and forgetting
all about its conjugate variable $T$. This should not hide the
fundamentally different ways in which the Dirac--ADM action is obtained
from the spacetime action. The Lagrangian (9.2)--(9.3) for ordinary
dust is nondegenerate in the velocities $\dot{T}$, $\dot{Z}^k$. The
expressions for the momenta $P$, $P_k$ can be inverted to yield the
velocities. The momenta are in a one--to--one correspondence with the
multipliers $M$ and $W_k$ and hence their variation yields equivalent
equations. The spacetime action, so to speak, is in an `already
parametrized form'.

To cast the spacetime action (4.4)--(4.5) of the null dust into
canonical form requires an entirely different procedure. The null dust
Lagrangian (6.8), (6.10) is singular in the velocities
$\dot{Z}^k$. The definition equations for the momenta $P_k$ cannot be
inverted. They yield three constraints
\begin{equation}
\delta^{ijk}W_j P_k =0
\end{equation}
demanding that the multipliers $W_k$ be parallel to the momenta
$P_k\,$, which leaves the magnitude of $W_k$ undetermined. The
variation of the action with respect to $W_k$ leads to the constraint
(6.14) on the velocities $\dot{Z}^k$. If this constraint is satisfied,
the multipliers $W_k$ can be replaced by a single multiplier $W$ and
the Lagrangian $L^{\rm ND}$ cast into an equivalent form (6.16) which
is regular in the velocities. This allows one to perform the Legendre
dual transformation to the canonical form of the action. The final
elimination of the multiplier $W$ (analogous to the final elimination
of the mass multiplier $M$ from the canonical action for ordinary
dust) leads to the null dust momentum and energy densities (6.22) and
(6.28). To summarize, though these final expressions have similar
structure as the densities (9.9)--(9.10) for ordinary dust from which
they can be obtained by putting $P=0$, their derivation is
fundamentally different.

After the dust is coupled to geometry, the parallels and differences
between ordinary and null dust are brought into a new perspective. The
momentum and Hamiltonian  constraints for ordinary dust can be resolved
with respect to the four dust momenta $P$, $P_k$ which brings them to
an equivalent form
\begin {eqnarray}
H_{\uparrow k} \,&\mbox{:=}&\, P_k + Z^a_k H^{\rm G}_{a} +
\sqrt{G}\; T_{,a} Z^a_k =0\,,\\
H_{\uparrow}\, &\mbox{:=}&\, P - \sqrt{G} = 0
\,,\end{eqnarray}
where $G$ is given by Eq. (7.2). The new constraint functions
$H_{\uparrow K} = ( H_{\uparrow}\,,H_{\uparrow k})$ have strongly
vanishing Poisson brackets:
\begin{equation}
\left\{ H_{\uparrow K}(x)\,,\, H_{\uparrow L}(x') \right\}=0\,.
\end{equation}

The imposition of the constraints (9.13) and (9.14) as operator
restrictions on the states $\Psi [T, Z^{k};\,g_{ab}\,,p^{ab}]$ leads
to a functional Schr\"{o}dinger equation with formally conserved inner
product. By mapping the constraints into the dust space, the momentum
constraint is eliminated and what remains is a single functional
differential Schr\"{o}dinger equation
\begin{equation}
\left( \widehat{\bbox{P}}(z) - \sqrt{{\bbox{G}}(z;\,\widehat
{\bbox{g}},\,\widehat{\bbox{p}}}]
\right) {\bbox{\Psi}}[{\bbox{T}}(z), {\bbox{g}}(z)]=0\,.
\end{equation}

The null dust constraints in the form (8.6), (7.2) can again be
obtained from the ordinary dust constraints (9.13)--(9.14) by
disregarding the canonical pair $T$, $P$ (and squaring Eq. (9.14)). By
mapping them into dust space, the momentum constraint is again
eliminated. By imposing the only remaining constraint as an operator
restriction on quantum states, one again gets a single functional
differential equation (8.18). However, and this is an important
difference, Eq. (8.18) is {\em not} a Schr\"{o}dinger equation like
Eq. (9.16) because there is no internal time ${\bbox{T}}(z)$. It is
thus not clear how to introduce an inner product in the space of its
solutions.

Both ordinary dust and null dust provide a standard of space in
canonical gravity because the dust particles introduce into spacetime
a privileged dust frame $\cal S$ labeled by comoving coordinates
$Z^k(x)$. The crucial difference is that ordinary dust provides also a
standard of time: It has an additional degree of freedom $T(x)$ which
can be physically interpreted as the proper time along the dust
worldlines. Null dust does not have any corresponding degree of
freedom because affine parametrization of null geodesics is
ambiguous. It thus fails to provide a standard of time to the
spacetime in which it moves. The story of ordinary dust is that of
time regained. The story of null dust is that of time lost again.

\section{Acknowledgments}

We are grateful to Laszlo Gergely and Jorma Louko for their comments.
The work on this paper was partially supported by the U.S.--Czech
Republic Science and Technology Program grant 92067, by the NSF
grant PHY--9507719 to the University of Utah, and by the Czech
Republic grant GACR--202/96/0206.


\appendix
\section{Null dust and geometrical optics}

If at each spacetime point all the energy is transported in one
direction with the speed of light, it is appropriate to describe
the matter by the energy--momentum tensor of null dust,
\begin{equation}
T^{\alpha \beta}= M k^\alpha k^\beta\,.\eqnum{A1}
\end{equation}
The energy--momentum tensor (A1) may be considered as representing an
incoherent superposition of waves with random phases and polarizations
but moving in a single direction.\footnote[6]{This is different from
the energy--momentum tensor of perfect fluid with the equation of
state $p = M/3$, which represents the superposition of waves with
random propagation directions.} It is also called the
`geometrical--optics' or `pure radiation' energy--momentum tensor.

As an example, consider the Maxwell theory. (See, in
particular, \cite{MTW}, \S 22.5, for a detailed exposition of
geometrical optics in curved spacetime.) If the electromagnetic
waves can {\it locally} be regarded as plane waves
propagating through spacetime of negligible curvature, one can write
the electromagnetic vector potential $A_\alpha$ in the form
\begin{equation}
A_\alpha = {\rm Re} \left( a_\alpha e^{i\Theta} \right)\,.\eqnum{A2}
\end{equation}
Here, in the first approximation, the complex amplitude
$a_{\alpha}(y)$ is independent of the wavelength and is slowly
changing as a function of spacetime position $y$, while the scalar
function $\Theta (y)$ is a rapidly changing phase. Following the
standard procedure \cite{MTW}, one introduces the wave vector
\begin{equation}
k_\alpha = \Theta _{,\alpha}\,,\eqnum{A3}
\end{equation}
the (real) scalar amplitude
\begin{equation}
A = (A_\alpha A^\alpha)^{1/2} = (a^\alpha \bar
{a}_\alpha)^{1/2},\eqnum{A4}
\end{equation}
and the (complex) unit polarization vector
\begin{equation}
e_\alpha = A^{-1} a_\alpha\,.\eqnum{A5}
\end{equation}
As a consequence of the source--free wave equation and the Lorentz
gauge condition, both written in the first order of the
geometrical optics approximation, the quantities (A3)--(A5)
obey the following set of equations:
\begin{eqnarray}
k_\alpha k^\alpha &=& 0\,,\eqnum{A6}\\
k^\beta \nabla_\beta k^\alpha &=& 0\,,\eqnum{A7}\\
\nabla_\alpha ( A^2 k^\alpha) &=& 0\,,\eqnum{A8}
\end{eqnarray}
and
\begin{equation}
k^\alpha e_\alpha = 0\,,\qquad  k^\beta \nabla_\beta e_\alpha = 0\,.
\eqnum{A9}
\end{equation}
From Eq. (A7) we see that the null vector $k^\alpha$ is affinely
parametrized. The electromagnetic field tensor is given by
\begin{equation}
F_{\alpha \beta} = 2 {\rm Re} \left( i A e^{i {\Theta}} k_{[\alpha}
e_{\beta]} \right).\eqnum{A10}
\end{equation}
It represents the electromagnetic field of type $ N$ (the null
field) since it satisfies the relations
\begin{equation}
(F_{\alpha \beta} + i F^*_{\alpha \beta}) k^\beta = 0\,, \;\;\;\;\;
F_{\alpha \beta} F^{\alpha \beta} = F_{\alpha \beta} F^{*\alpha \beta} 
=0\,,\eqnum{A11}
\end{equation}
where $F^*_{\alpha \beta}$ is dual to $F_{\alpha \beta}\,$.
Equations (A7) and (A8) imply the covariant conservation law
for the electromagnetic energy--momentum tensor
\begin{equation}
T^{\alpha \beta} = A^2 k^\alpha k^\beta .\eqnum{A12}
\end{equation}

We see that the phenomenological null dust equations (2.9),
(3.16)--(3.18) are the same as Eqs. (A6)--(A8) and (A12) of the
high--frequency limit of the Maxwell theory if the null vector field
$k_\alpha$ is defined by Eq. (A3) and the mass distribution $M$ is
identified with the square of the scalar amplitude $ A\,$:
\begin{equation}
M= A^2.\eqnum{A13}
\end{equation}

Null dust thus  exhibits all features of the geometrical optics
limit of Maxwell's theory except for the polarization properties.
However, starting from a solution of the null dust equations one
can always construct a polarization vector $e_\alpha$ such that
Eqs. (A9) are also satisfied. This yields the tensor
(A10) which can be regarded as an electromagnetic field tensor
in the geometrical optics approximation.

The laws of geometrical optics can also be interpreted as describing
photons that move along null rays with the flux vector which is
determined by the amplitude $A$ and the null vector $k^\alpha$
(see \cite{MTW} for details).

The lightlike particles need not necessarily be photons. It is
quite obvious that similar conclusions can be reached for all
zero--rest--mass fields in high--frequency limit. For example, by
employing the geometrical optics form (A1) of the
energy--momentum tensor, several authors \cite{col-neu} studied the
gravitational collapse with escaping neutrinos.

A somewhat special case is the gravitational field itself.  Careful
studies of the high--frequency limit of the gravitational radiation by
Isaacson and others \cite{isaa} have shown that the energy--momentum
tensor (A12) and the null vector field $k^\alpha$ which satisfy
Eqs. (A6)--(A8) also describe the behavior of high--frequency
gravitational waves. The metric tensor perturbations representing
high--frequency waves are given by
\begin{eqnarray}
h_{\alpha \beta} & = & {\rm Re} \left( (a_{\alpha \beta} -
{\textstyle{1\over2}} a \stackrel{_0}{\gamma}_{\alpha \beta})e^{i
\Theta}\right)\,, \nonumber \\ a & := & \stackrel{_0}{\gamma}^{\alpha
\beta}a_{\alpha \beta}\,,\eqnum{A14}
\end{eqnarray}
where $\stackrel{_0}{\gamma}_{\alpha \beta}$ is the background metric
(the source of which may be the high--frequency waves themselves).
By applying the geometrical optics approximation to the perturbed
Einstein's equations, one arrives again at the equations (A3),
(A6)--(A8), and (A12). Instead of the scalar amplitude (A4) one now
gets
\begin{equation}
A=\left({\textstyle{1\over2}} a^{\alpha \beta} \bar {a}_{\alpha
\beta}\right)^{1/2}.\eqnum{A15}
\end{equation}
One also obtains the equations for the polarization tensor $e_{\alpha
\beta}=a_{\alpha\beta}/A$, analogous to Eqs. (A9) (see \cite{MTW}, 
\cite{isaa}). The Riemann tensor of the metric (A14) has the Petrov
type $N$. The gravitational field in the high--frequency limit is null,
similarly as the electromagnetic field. The well--known peeling--off
property of exact radiative (zero--rest--mass) fields in asymptotically
flat spacetimes \cite{sachs} implies that at large distances from the
source these fields are null, having the structure of plane waves. In
asymptotic regions one can even describe {\it exact} solutions of
the field equations in terms of null dust. In such situations, one can
usually find a natural parametrization of null rays -- for example,
by the proper time of distant observers at rest with respect to an
isolated source.

The variational approach of MacCallum and Taub \cite{mc+t} to the
high--frequency gravitational waves is especially relevant for the
present paper. By applying the `averaged Lagrangian technique' of
Witham to the second variation Lagrangian for the perturbations of
vacuum gravitational field, these authors give a variational principle
for approximately periodic gravitational wave described by metric
perturbation of the form (A14). Their principle, derived by perturbing
and averaging the Hilbert action, implies the geometrical optics
equations (A3), (A6)--(A8), and (A12), with $ A$ given by
Eq. (A15). This principle is closely related to our variational
principle for null dust (given in Eqs. (4.4), (4.5)), in the special
case of the hypersurface orthogonal vector field $l^{\alpha}$.
\section{Exact solutions with null dust: examples and some recent
applications} 

As an illustration, we shall give a few examples of known exact
spacetimes with null dust. (A detailed survey of such solutions found
before 1980 is given in \cite{kretal}. The cosmological solutions with
null dust were recently reviewed in \cite{kras}, and the solution
representing colliding plane gravitational waves accompanied by null
dust in \cite{grif}.)

Among the simplest solutions directly related to the fields arising in
the geometrical optics limit are conformally flat null dust solutions
representing special plane waves. They are described by the line
element (see, e.g., \cite {kretal})
\begin{equation}
ds^2 = -\frac{1}{4} \Phi^2(u_{-})(x^2+y^2)du^2_{-}
-2du_{+}du_{-} +dx^2+dy^2\,,
\eqnum{B1}
\end{equation}
where $\Phi$ is an arbitrary function of a retarded time $u_{-}\,$. The
corresponding energy--momentum tensor is
\begin{equation}
T_{\alpha \beta} = \Phi^2k_\alpha k_\beta\,;\eqnum{B2}
\end{equation}
the only nonvanishing component of the null covector $k_\alpha$ is
$k_{{u}_{-}} = 1$. These solutions can always be interpreted as exact
solutions of the Einstein--Maxwell equations with the null
electromagnetic field given by $F_{\alpha \beta} =
2\Phi(u_{-})k_{[\alpha}e_{\beta]}$, where $e_\alpha = (0,\,0,\,\cos
\psi,\, \sin \psi)$ contains an arbitrary function $\psi =\
\psi(u_{-})$ (cf. Eq.  (A10)). Cylindrical gravitational waves
accompanied by null dust are also known \cite{rao}.

A more complicated class of radiative solutions with `spherical'
gravitational waves and null dust is formed by the Robinson--Trautman
solutions \cite{traut}.  The energy--momentum tensor has again the
form (B2), but the function $\Phi$ is now given by $\Phi^2 = {\em
n}^2(\zeta, \bar{\zeta}, u_{-})/v^2$, where $\zeta$ is a complex
spatial coordinate, $v$ is an affine parameter along the rays, and
$u_{-}$ is a retarded time. The function ${\em n}$ may be
arbitrary. If, however, these solutions should represent exact
Einstein--Maxwell fields, ${\em n}$ must have the form $
n^2=2h\bar{h}P^2$, where $h(\zeta, \bar{\zeta}, u_{-})$ and $P(\zeta,
\bar{\zeta}, u_{-})$ satisfy certain additional conditions \cite{kretal}. 
The Robinson--Trautman solutions with null dust include Vaidya's
spherically symmetric metric as a special case. In fact, if the
evolving null dust is homogeneous, all such Robinson--Trautman
spacetimes approach the Vaidya's metric as the retarded time goes to
infinity \cite{bp}.

The null vector field $k^\alpha$ in the solutions we have mentioned is
hypersurface orthogonal and the corresponding null congruence is thus
nontwisting. The twisting null dust solutions are discussed in
\cite{her}, the best known simple example being the `radiating Kerr
metric'.

Some exact solutions with null dust can also be interpreted as exact
solutions of Einstein's equations coupled to a massless scalar field
\cite{scal}. However, given a conserved energy--momentum tensor in the
form (A1), it is not necessarily true that the mass distribution $M$
and the null vector field $k^{\alpha}$ represent an electromagnetic or
a massless scalar field. However, if the null vector field $k^\alpha$
is shear--free, a corresponding nontrivial solution of Maxwell's
equations can be found by virtue of the Mariot--Robinson theorem
\cite{rob}.

Recently, certain exact solutions with null dust which can be
interpreted as `relativistic rockets' have been explored in connection
with the properties of gravitational radiation \cite{bonn}. A number of
studies have also been devoted to colliding plane and cylindrical
systems with null dust \cite {col}. 

Above all, as we have already stated in the Introduction, the null
dust models have been recently used to clarify the formation of naked
singularities during a spherical gravitational collapse \cite{naks},
in the studies of the mass inflation inside black holes \cite{minfl},
and in the models attempting to describe the formation and Hawking
evaporation of black holes
\cite{haev}.
\section{Description of twisting null congruences
by Pfaff forms: Two examples}

Since null congruences are somewhat unusual, we give here two examples
of twisting null congruences described by the scalar potentials $Z^i$
and $w_i$.

{\bf I.}  In a flat spacetime with Lorentzian coordinates $(t, x, y,
z)$, consider a system of lightlike particles which, in each plane
perpendicular to the $z$--axis, move in mutually parallel straight
lines. As one passes from one plane $z=const$ to another, the angle
$\alpha$ between particle trajectories and the $x$--axis smoothly
changes with $z:$ $\alpha=\alpha(z) \in [0,2\pi)$. It is easy to see
that the null worldlines form a twisting null congruence:
\begin{eqnarray}
t & = & v+t_0\,, \nonumber \\
x & = & v\, \cos \alpha(z) + x_0\,, \eqnum{C1}\\
y & = & v\,\sin \alpha(z) + y_0\,, \nonumber \\
z & = & z_0\,,\nonumber
\end{eqnarray}
where $v \in {\hbox{$I$\kern-3.8pt $R$}}$ is an affine parameter. The
tangent null vectors $k^\alpha=dx^\alpha/dv$ are given by
\begin{equation}
k^{\alpha}=(1,\, \cos \alpha(z),\, \sin \alpha(z),\, 0)\,. \eqnum{C2}
\end{equation}
One can readily check that
\begin{eqnarray}
k^{\alpha}k_{\alpha}
=0\,, \;\;\;\;\;\;\; k^{\beta}\nabla_{\beta}k_{\alpha}=0\,,
 \eqnum{C3}
\end{eqnarray}
confirming that (C1) is a congruence of null geodesics affinely
parametrized by $v$.

The first comoving coordinate
\begin{equation}
Z^1 = z  \eqnum{C4}
\end{equation}
is trivial: It determines the plane in which the geodesic lies. The
second comoving coordinate $Z^2$ is the coordinate $y'$ of the
cartesian system $(x', y', z)$ obtained from $(x, y, z)$ by the
rotation about the $z$--axis by the angle $\alpha(z)$:
\begin{equation}
Z^2 = - x \sin \alpha(z) + y \cos \alpha(z)\,. \eqnum{C5}
\end{equation}
In the rotated cartesian systems $(x', y', z)$, the particles move
along the $x'$--axes, with $y'=const$. The third comoving
coordinate $Z^3$ is the retarded time $u_-$ corresponding to that
direction:
\begin{eqnarray}
Z^3=u_{-}=t-x' = t-x \cos \alpha(z) -y \sin \alpha(z)\,. \eqnum{C6}
\end{eqnarray}
From Eqs. (C4)--(C6) we obtain the covectors ${Z^k}_{,\alpha\,}$:
\begin{eqnarray}
{Z^1}_{,\alpha} & = & {\bf (} 0,\,0,\,0,\,1{\bf )}\,, \nonumber \\
{Z^2}_{,\alpha} & = & {\bf (} 0,\,-\sin \alpha (z),\, \cos \alpha
(z),\, -x\alpha'(z) \cos \alpha (z) -y\alpha'(z) \sin \alpha(z){\bf )} \,,
\eqnum{C7} \\ {Z^3}_{,\alpha} & = & {\bf (} 1,\,-\cos \alpha(z),\,
-\sin \alpha (z),\, x\alpha'(z) \sin \alpha (z) - y \alpha'(z) \cos \alpha
(z){\bf )}\,, \nonumber
\end{eqnarray}
where $\alpha'\,\mbox{:=}\, d\alpha/dz$. It is easy to see that
${Z^k}_{,\alpha}$ are independent covectors. Since
\begin{equation}
k_{\alpha}={\bf (} -1,\,\cos \alpha(z),\, \sin \alpha(z),\,0{\bf )}\,,
\eqnum{C8}
\end{equation}
the decomposition (5.5) 
is obtained with the coefficients
\begin{equation}
w_1 = x \alpha'(z)\sin \alpha(z) - y\alpha'(z)\cos \alpha(z)\,,
\;\;\;w_2=0\,,\;\;\; w_3 = -1\,. \eqnum{C9}
\end{equation}
One can easily check that $w_k$ are constant along the geodesics,
Eq. (5.8).  This also follows from Eq. (C1) which allows us to write
$w_1$ in the form $w_1 = \alpha'(z_0)\,{\bf(} x_0\sin \alpha(z_0) - y_0
\cos \alpha (z_0){\bf )}\,$. Similarly, one can check that $k^\alpha
{Z^k}_{,\alpha}= 0\,$, as given by Eq. (4.6). One can also check the fact
mentioned in Section 4, that for a twisting congruence all vectors
${Z^k}_{,\alpha}$ are spacelike (except perhaps a set of measure
zero). The spacelike character of the vectors ${Z^1}_{,\alpha}$ and
${Z^2}_{,\alpha}$ is evident; for ${Z^3}_{,\alpha}$ we have
\begin{equation}
\eta^{\alpha\beta}{Z^3}_{,\alpha}{Z^3}_{,\beta}=\alpha'(z)^2\,
 {\bf (}x \sin \alpha(z) - y \cos\alpha(z){\bf )}^2. \eqnum{C10}
\end{equation}
The vector ${Z^3}_{,\alpha}$ is thus spacelike unless $\alpha' = 0$.
Calculating the twist $\omega$ of our congruence {\bf(}see Eq.
(3.24){\bf)}, we find
\begin{equation}
\omega = \frac{1}{2}|\alpha'|. \eqnum{C11}
\end{equation}
Hence, if the congruence is twisting, all the three vectors
${Z^k}_{,\alpha}$ are spacelike. When $\alpha' = 0$, $k_\alpha =
{Z^3}_{,\alpha}\,,\,$ so that the congruence is hypersurface orthogonal.

Instead of the comoving coordinates $Z^k,\,$ one can, of course, use
other comoving variables $Z^{k'}= Z^{k'}(Z^i)$. Also, one can
parametrize the geodesics by a label $u$ different from the affine
parameter $v$. When we change the parameterization, $v = v(u, Z^k)$,
the null vectors are rescaled:
\begin{equation}
k^\alpha \rightarrow
U^\alpha = \frac{dx^\alpha}{d u} = \frac{\partial v}{\partial
u} k^\alpha . \eqnum{C12}
\end{equation}
This leads to a new decomposition, namely
\begin{equation}
U_{\alpha}={\rm W}_2 {Z^2}_{,\alpha}+{\rm W}_3 {Z^3}_{,\alpha}\,,
\eqnum{C13}
\end{equation}
where ${\rm W}_2=(\partial v/ \partial u) w_2\,$, ${\rm W}_3 =\partial
v/\partial u\,$.  As discussed in Section 5, if the congruence is not
affinely parametrized, i.e., if $\partial v/\partial u  \not=
const$, the coefficients ${\rm W}_k$ are not necessarily comoving.

{\bf II.}  The second example will be described only briefly. It is
the familiar ingoing principal null congruence in Kerr spacetime. In
the ingoing Kerr coordinates $(\widetilde {V}, r, \theta, \tilde
{\varphi})$ which generalize the ingoing Eddington--Finkelstein
coordinates of the Schwarzschild metric, the Kerr metric reads (our
notation follows \cite{MTW}):
\begin{eqnarray}
ds^2 = & -& (1-2Mr\rho^{-2})d\widetilde{V}^2 + 2dr d\widetilde{V} +\rho^2
d\theta^2 \nonumber
 \\& +& \rho^{-2}\left[ (r^2+a^2)^2- \Delta a^2 \sin^2 \theta
\right] \sin^2 \theta d\tilde{\varphi}^2 \eqnum{C14}
 \\ & - & 2a \sin^2 \theta
d\tilde{\varphi} dr -4aMr\rho^{-2}\sin^2
\theta d\tilde{\varphi} d\widetilde{V}\,.\nonumber
\end{eqnarray}
Here, the constant parameters $M$ and $a$ are the mass and angular
momentum per unit mass, and the functions $\Delta$ and $\rho$ have the
form
\begin{equation}
\Delta = r^2 - 2Mr + a^2,\;\;\;\;\;\; \rho^2 = r^2 + a^2 \cos^2 \theta.
 \eqnum{C15}
\end{equation}

The ingoing null Kerr congruence is given by
\begin{equation}
\widetilde{V} = const, \qquad    r = -v, \qquad
\theta = const, \qquad       \tilde{\varphi} = const, \eqnum{C16}
\end{equation}
where we have absorbed a constant energy parameter into the
affine parameter $v$ (cf. \cite{MTW}). The coordinates $Z^1 \,\mbox{:=}\,
\widetilde{V},\; Z^2 \,\mbox{:=}\, \theta,\; Z^3 \,\mbox{:=}\,
\tilde{\varphi}$ are clearly comoving. We can easily form the basis
vectors ${Z^k}_{,\alpha}\,$:
\begin{eqnarray}
{Z^1}_{,\alpha} & = & (1,\, 0,\, 0,\, 0)\,, \nonumber \\
{Z^2}_{,\alpha} & = & (0,\, 0,\, 1,\, 0)\,, \eqnum{C17}\\
{Z^3}_{,\alpha} & = & (0,\, 0,\, 0,\, 1)\,. \nonumber 
\end{eqnarray}
The covariant components of the tangent null vector $k^\alpha =
dx^\alpha/dv$ are
\begin{equation}
k_\alpha = (-1,\, 0,\, 0,\, a \sin^2 \theta)\,. \eqnum{C18}
\end{equation}
Its decomposition into the three covectors ${Z^k}_{,\alpha}$ yields
the coefficients
\begin{equation}
w_1 = -1\,,\;\;\; w_2 = 0\,,\;\;\; w_3 = a \sin^2 \theta\,, \eqnum{C19}
\end{equation}
which are constant along the null geodesics (C16), in accordance with
Eq. (5.8). The covariant metric can be read off from Eq. (C14).
The norms of the vectors ${Z^k}_{,\alpha}$ are
\begin{eqnarray}
& & g^{\alpha \beta}{Z^1}_{,\alpha} {Z^1},_\beta = \rho^{-2} a^2 \sin^2
\theta\,,\nonumber \\
& &g^{\alpha \beta} {Z^2}_{,\alpha} {Z^2}_{,\beta} = \rho^{-2},
\eqnum{C20} \\ & & g^{\alpha \beta} {Z^3},_\alpha {Z^3},_\beta = (\rho
\sin\theta)^{-2},
\nonumber
\end{eqnarray}
where $\rho^2$ is given by (C15). We see that all the vectors
${Z^k}_{,\alpha}$ are spacelike as long as $a\not= 0$, i.e., when the
congruence is twisting. The twist $\omega$, given by Eq. (3.24), is
\begin{equation}
\omega = |a \cos\theta| \rho^{-2}. \eqnum{C21}
\end{equation}

The congruence (C16) is twisting even in the flat--space limit of
the Kerr metric, obtained by putting $M=0$. In fact, Eqs. (C20)
and (C21) are independent of $M$. With $a = 0$, the vector
${Z^1}_{,\alpha}$ becomes null and the congruence is hypersurface
orthogonal, $k_\alpha = -{Z^1}_{,\alpha}\,$, i.e., nontwisting.

The comoving coordinates $Z^2 = \theta,\; Z^3 = \tilde{\varphi}$ are
simple, but they become singular at the axis $\theta = 0$ and $\theta
= \pi\,$, the magnitude of the vector ${Z^3}_{,\alpha}$ becoming
infinite, Eq. (C20). It is easy, however, to cure this defect by
going over to another pair of comoving coordinates, $Z^{2'}$ and
$Z^{3'}$, e.g.,
\begin{equation}
Z^{2'} = \sin\tilde{\varphi}\, \sin\theta,\;\;\;Z^{3'}=
\cos\tilde{\varphi}\, \sin\theta\,. \eqnum{C22}
\end{equation}
Then
\begin{equation}
k_\alpha = - {Z^1}_{,\alpha} + a\sin\theta
\cos\tilde{\varphi}\,{Z^{2'}}_{,\alpha} - a \sin \theta \sin
\tilde{\varphi} \,{Z^{3'}}_{,\alpha}\,, \eqnum{C23}
\end{equation}
and
\begin{eqnarray}
g^{\alpha \beta} {Z^{2'}}_{,\alpha} {Z^{2'}}_{,\beta} & = &
\rho^{-2} (1 - \sin^2 \tilde{\varphi}\, \sin^2 \theta)\,,\nonumber \\
g^{\alpha \beta} {Z^{3'}}_{,\alpha}{Z^{3'}}_{,\beta} & = &
\rho^{-2} (1 - \cos^2 \tilde{\varphi}\, \sin^2 \theta)  \eqnum{C24}
\end{eqnarray}
are regular at $\theta = 0$ and $\theta = \pi$.
\nopagebreak[4]


\begin{thebibliography}{99}


\bibitem{B+K} J. D. Brown and K. V. Kucha\v{r}, Phys. Rev. D {\bf
  51}, 5600 (1995). Brief accounts of this work are K. V. Kucha\v{r},
  in {\em Directions in General Relativity I}, edited by
  B. L. Hu, M. P. Ryan, and C. V. Vishveshwara (Cambridge University
  Press, Cambridge, 1993), and J. D. Brown, in {\em Proceedings of the
  Cornelius Lanczos Centenary Conference}, edited by J. D. Brown,
  M. T. Chu, D. C. Ellison, and R. J. Plemmons (SIAM, Philadelphia,
  1994).

\bibitem{vai} P. C. Vaidya, Nature {\bf 171}, 260 (1953).
  Here the metric is given in terms of a retarded time coordinate
  (called by Vaidya a Newtonian time). Originally, Vaidya found
  the metric in the Schwarzschild--type coordinates in which it has
  a more complicated form; see P. C. Vaidya, Curr. Sci. {\bf 12},
  183 (1943), and the detailed version: P. C. Vaidya, Proc.
  Indian Acad. Sci. {\bf A33}, 264 (1951).

\bibitem{MTW} C. W. Misner, K. S. Thorne, and J. A. Wheeler, {\em
  Gravitation\/} (Freeman, San Francisco, 1973).

\bibitem{sachs} R. Sachs, Proc. Roy. Soc. London A {\bf 264}, 309
  (1961). For a review, see F. A. E. Pirani, in {\em Lectures on
  general relativity}, edited by S. Deser and K. W. Ford (Prentice
  Hall, Englewood Cliffs, 1965).

\bibitem{P+R} R. Penrose  and W. Rindler, {\em Spinors and
  space--time II} (Cambridge University Press, Cambridge, 1986).

\bibitem{bmb} H. Bondi, M. G. J. van der Burg, and A. W. K. Metzner,
  Proc. Roy. Soc. London A {\bf 269}, 21 (1962).

\bibitem{kr+sa} J. Kristian and R. K. Sachs, Astrophys. J. {\bf
  143}, 379 (1966).

\bibitem{cara} C. Carath\'{e}odory, {\em Variationsrechnung und
  Partielle Differentialgleichungen Erster Ordnung I} (Teubner,
  Leipzig, 1956). Pfaff's theorem is well described, for example,
  in Appendix A of B. Schutz, Phys. Rev. D {\bf 2}, 2762
  (1970).

\bibitem{FluidS} Our approach follows \cite{B+K}. The general form
  of the action is patterned after the relativistic perfect fluid
  action introduced in J. D. Brown, Class.  Quantum Grav. {\bf 10},
  1579 (1993). This action makes use of a velocity--potential or
  `Clebsch' representation of the fluid velocity (A. Clebsch, J. Reine
  Angew. Math. {\bf 56}, 1 (1859)).  Clebsch potentials were used in
  the development of action principles for nonrelativistic fluids by
  Lin (C. C. Lin, in {\em Liquid Helium}, edited by G. Careri
  (Academic Press, New York, 1963); see also J. Serrin, in {\em
  Handbuch der Physik}, volume 8, edited by S. Fl{\"u}gge and
  C. Truesdell (Springer, Berlin, 1959)) and by Seliger and Whitham
  (R. L. Seliger and G. B.  Whitham, Proc. Roy. Soc. A {\bf 305}, 1
  (1968)). Tam used Clebsch potentials to formulate an action
  principle for an ideal, charged fluid in the context of special
  relativity (K. Tam, Can. J. Phys.  {\bf 44}, 2403 (1966)). Action
  principles for perfect fluids in general relativity were developed
  independently by Tam and O'Hanlon (K. Tam and J. O'Hanlon, Nuovo
  Cimento {\bf 62}B, 351 (1969)) and by Schutz {\bf (} B. F. Schutz,
  Phys. Rev. D {\bf 2}, 2762 (1970)\,{\bf )}. The actions discussed
  by Tam and O'Hanlon and by Schutz use a minimal set of velocity
  potentials, which precludes the interpretation of the appropriate
  potential fields as Lagrangian coordinates for the fluid. The
  Hamiltonian form of Schutz's action was developed by Schutz
  (B. F. Schutz, Phys.  Rev. D {\bf 4}, 3559 (1971)) and by Demaret
  and Moncrief (J. Demaret and V. Moncrief, Phys. Rev. D {\bf 21},
  2785 (1980)).

\bibitem{kk-kh} K. V. Kucha\v{r}, J. Math. Phys. {\bf 17}, 792 (1976).

\bibitem{K+R} K. V. Kucha\v{r} and J. D. Romano, Phys. Rev. D {\bf 51},
  5579 (1995).

\bibitem{B+M} J. D. Brown and D. Marolf, Phys. Rev. D {\bf 53}, 1835 
  (1996).

\bibitem{M} F. Markopoulou, Class. Quantum Grav. {\bf 13}, 2577 (1996).

\bibitem{rainich} G. Y. Rainich, Trans. Am. Math. Soc. {\bf 27},
  106 (1925).

\bibitem{M+W} C. W. Misner and J. A. Wheeler, Ann. Phys. (NY) {\bf 2},
  525 (1957).

\bibitem{hlav} V. Hlavat\'{y}, Journ. de Math. {\bf 40}, 1 (1961).

\bibitem{per} A. Peres, Bull. Res. Counc. Israel {\bf 9}F, 129 (1960).

\bibitem{kk-rain} K. V. Kucha\v{r}, Czech. J. Phys. B {\bf 13},
  551 (1963).

\bibitem{kk-neu} K. V. Kucha\v{r}, Acta Phys. Pol. {\bf 28}, 695 (1965).

\bibitem{bica} J. Bi\v{c}\'{a}k, Czech. J. Phys. B {\bf 16}, 95 (1966).

\bibitem{col-neu} C. W. Misner, Phys. Rev. {\bf 137}, B1360 (1965);
  K. P. Singh and M. C. Srivastava, Proc. Natl. Inst. Sci. India A {\bf
  35}, 619 (1969); J. B. Griffiths and R. A. Newing, Gen.
  Rel. Grav. {\bf 5}, 345 (1974). The form (A1) of the energy--momentum
  tensor to describe the geometrical optics limit of the neutrino
  field equations was justified by J. B. Griffiths, Gen. Rel. Grav.
  {\bf 4}, 361 (1973); and references therein.

\bibitem{isaa} R. A. Isaacson, Phys. Rev. {\bf 166}, 1263 and
  1272 (1968); Y. Choquet--Bruhat, Commun. Math. Phys. {\bf 12}, 16
  (1969). See also \cite{MTW} and \cite{mc+t}. Interacting gravitational,
  electromagnetic and charged scalar fields in the high--frequency limit
  are discussed by Y. Choquet--Bruhat and A.H . Taub, Gen.
  Rel. Grav. {\bf 8}, 561 (1977). A precise characterization of the
  high--frequency limit in general relativity was given by C. A. Burnett,
  J. Math. Phys. {\bf 30}, 90 (1989). Burnett proved that there
  exists an effective energy--momentum tensor (which, in special cases,
  has the form (A1) appropriate for null dust) that describes
  high--frequency waves acting as a source of smooth background
  curvature. He did not need to assume hypersurface orthogonality of
  $k^\alpha$.

\bibitem{mc+t} M. A. H. MacCallum and A. H. Taub, Commun. Math.
  Phys. {\bf 30}, 153 (1973); for a review, see A. H. Taub, in {\it
  General Relativity and Gravitation}, edited by A. Held, (Plenum Press,
  New York, 1979). Whitham's technique averages over a rapidly changing
  phase directly in a Lagrangian. One thus averages a scalar density
  (the Hilbert Lagrangian) rather than tensor fields, as in the
  Brill--Hartle averaging used by Isaacson \cite{isaa}.

\bibitem{kretal} D. Kramer, H. Stephani, M. A. H. MacCallum, and E.
  Herlt, {\em Exact Solutions of Einstein's Field Equations}
  (Cambridge University Press, Cambridge, 1980).

\bibitem{kras} A. Krasi\'{n}ski, {\em Physics in an Inhomogeneous
  Universe}, preprint (Copernicus Astronomical Center, Warszaw 1993).
 
\bibitem{grif} J. B. Griffiths, {\em Colliding Plane Waves in
  General Relativity} (Oxford University Press, Oxford, 1991).

\bibitem{rao} J. Krishna Rao, Indian J. Pure Appl. Math.
  {\bf 1}, 367 (1970), and \cite{kretal}. There are only a few cases in
  which these solutions can be interpreted as exact solutions of the
  Einstein--Maxwell equations.

\bibitem{traut} I. Robinson and A. Trautman, Proc. Roy. Soc.
  London A {\bf 265}, 463 (1962).  
  A clear and concise discussion of Robinson--Trautman spacetimes with
  null dust is given in \cite{kretal}, \S 24.3.

\bibitem{bp} J. Bi\v{c}\'{a}k and Z. Perj\'{e}s, Class.
  Quantum Grav. {\bf 4}, 595 (1987).

\bibitem{her} E. Herlt, Gen. Rel. Grav. {\bf 12}, 1 (1980).

\bibitem{scal} Indeed, assuming that the scalar field satisfies the
  constraint $\phi_{,\alpha} \phi^{,\alpha} = 0\,$, its
  energy--momentum tensor assumes the form (A1) after the rescaling
  $k^\alpha \rightarrow \bar{k}^\alpha = M^{1/2} k^\alpha\,$. One can
  easily see that the condition $\phi,_\alpha \phi^{,\alpha} = 0$
  (which is always satisfied in the geometrical optics approximation)
  holds exactly for the combined gravitational and massless--scalar
  plane waves. See W. Z. Chao, J. Phys. A: Math. Gen. {\bf 15}, 2429
  (1982), and M. Halilsoy, Lett. Nuovo Cim. {\bf 44}, 544 (1985), who
  considered the collision of such waves.

\bibitem{rob} I. Robinson, J. Math. Phys. {\bf 2}, 290
  (1961). See also \cite{kretal}.

\bibitem{bonn} W. B. Bonnor, Class. Quantum Grav. {\bf 11}, 2007
  (1994); T. Damour, Class. Quantum Grav. {\bf 12}, 725
  (1995).

\bibitem{col} These works are mostly related to colliding
  plane gravitational waves \cite{grif}. Reference \cite{grif} also
  discusses the problem (widely explored in the years 1986--91) that
  colliding plane waves accompanied by null dust do not have a unique
  evolution. The problem was introduced by the work of S. Chandrasekhar
  and C. Xanthopolous, Proc. Roy. Soc. London A {\bf 403}, 189
  (1986), and its discussion led to an understanding that the
  time evolution of null dust may imply a `transmutation of matter'
  (here into a stiff fluid) if one does not distinguish different types
  of null dust representing different types of matter fields. For a
  recent work on the collision and interaction of cylindrically
  symmetric clouds of null dust, see P. S. Letelier and A. Wang, 
  Phys. Rev. D {\bf 49}, 5105 (1994).

\bibitem{naks} Y. Kuroda, Prog. Theor. Phys. {\bf 72}, 63
  (1984), and A. Papapetrou in {\em A random walk in relativity and
  cosmology}, edited by J. Krishna-Rao (Wiley Eastern, New Delhi, 1985)
  were the first who noticed that the collapse of spherical shell of
  null dust may lead to a naked singularity. A number of papers have
  then addressed this problem -- see, e.g., I. H. Dwivedi and
  P. S. Joshi, Class. Quantum Grav. {\bf 6}, 1599 (1989), J. S. Lemos,
  Phys. Rev. Lett. {\bf 68}, 1447 (1992), and references
  therein.

\bibitem{minfl} See, e.g., E. Poisson and W. Israel, Phys. Rev. D 
  {\bf 41}, 1796 (1990), R. Balbinot and E. Poisson,  Phys.
  Rev. Lett. {\bf 70}, 13 (1993), W. G. Anderson, P. R. Brady, W.
  Israel, and S. M. Morsink, Phys. Rev. Lett. {\bf 70}, 1041
  (1993), W. Israel, Int. J. Modern Physics D {\bf 3}, 71
  (1994); D. A. Konkowski and T. M. Helliwell,  Phys. Rev. D {\bf
  50}, 841 (1994); see also references therein.

\bibitem{haev} R. Parentani and T. Piran, Phys. Rev. Lett.
  {\bf 73}, 2805 (1994).

\end{thebibliography}
\end{document}